\documentclass[prd,amsmath,notitlepage,twocolumn,10pt]{revtex4-2}
\usepackage{graphicx}
\usepackage{float}
\usepackage{epstopdf,cancel}
\usepackage{epsf,latexsym,bbm,euscript}
\usepackage{amssymb,amsmath}
\usepackage{mathtools} 
\usepackage{times,graphics}
\usepackage{mathrsfs}
\usepackage{soul,xcolor}
\usepackage{mathtools}
\usepackage{braket}
\usepackage[normalem]{ulem}
\usepackage{setspace}
\usepackage{url,hyperref}
\hypersetup{colorlinks,linkcolor={blue!55!black},citecolor={red!45!black},urlcolor={blue!45!black},breaklinks=true}
\usepackage{tensor}
\usepackage[caption=false]{subfig}
\newcommand{\subfigimg}[3][,]{%
	\setbox1=\hbox{\includegraphics[#1]{#3}}%
	\leavevmode\rlap{\usebox1}%
	\rlap{\hspace*{-18pt}\raisebox{.5\baselineskip}{\small{#2}}}%
	\phantom{\usebox1}%
}

\def\sg{\textsl{g}}
\def\6{{\langle}}
\def\9{{\rangle}}
\def\pad{\partial}
\def\cF{\mathcal{F}}
\def\cO{\mathcal{O}}

\newcommand\ContFracOp{%
	\operatornamewithlimits{%
		\mathchoice
		{\vcenter{\hbox{\Large $\mathbf{K}$}}}
		{\vcenter{\hbox{ $\mathbf{K}$}}}
		{\mathbf{K}}
		{\mathbf{K}}}}

\usepackage{mathtools}
\newcommand{\defeq}{{\vcentcolon=}}
\newcommand{\eqdef}{{=\vcentcolon}}

\newcommand{\be}{\begin{equation}}
	\newcommand{\ee}{\end{equation}}
\newcommand{\p}{\partial}

\usepackage{scalerel}
\usepackage{tikz}
\usetikzlibrary{svg.path}
\definecolor{orcidlogocol}{HTML}{A6CE39}
\tikzset{
	orcidlogo/.pic={
		\fill[orcidlogocol] svg{M256,128c0,70.7-57.3,128-128,128C57.3,256,0,198.7,0,128C0,57.3,57.3,0,128,0C198.7,0,256,57.3,256,128z};
		\fill[white] svg{M86.3,186.2H70.9V79.1h15.4v48.4V186.2z}
		svg{M108.9,79.1h41.6c39.6,0,57,28.3,57,53.6c0,27.5-21.5,53.6-56.8,53.6h-41.8V79.1z M124.3,172.4h24.5c34.9,0,42.9-26.5,42.9-39.7c0-21.5-13.7-39.7-43.7-39.7h-23.7V172.4z}
		svg{M88.7,56.8c0,5.5-4.5,10.1-10.1,10.1c-5.6,0-10.1-4.6-10.1-10.1c0-5.6,4.5-10.1,10.1-10.1C84.2,46.7,88.7,51.3,88.7,56.8z};
	}
}
\newcommand\orcidlink[1]{\href{https://orcid.org/#1}{\mbox{\scalerel*{
				\begin{tikzpicture}[yscale=-1,transform shape]
					\pic{orcidlogo};
			\end{tikzpicture}}{X}}}}

\begin{document}
	
	\title{ Pad\'{e} metrics for black hole perturbations and light rings}
	
	\author{Swayamsiddha Maharana\,\orcidlink{0009-0004-6006-8637}}
	\email{swayamsiddha.maharana@hdr.mq.edu.au}
	
	\author{Fil Simovic\,\orcidlink{0000-0003-1736-8779}}
	\email{fil.simovic@mq.edu.au}
	
	\author{Ioannis Soranidis\,\orcidlink{0000-0002-8652-9874}}
	\email{ioannis.soranidis@hdr.mq.edu.au}
	
	\author{Daniel R. Terno\,\orcidlink{0000-0002-0779-0100}}
	\email{daniel.terno@mq.edu.au}
	
	\affiliation{School of Mathematical and Physical Sciences, Macquarie University, NSW 2109, Australia}

	\begin{abstract}
		\vspace*{1mm}

		Most distinguishing features of black holes and their mimickers are concentrated near the horizon. In contrast, astrophysical observations and theoretical considerations primarily constrain the far-field geometry. In this work we develop tools to effectively describe both, using the two-point Padé approximation to construct interpolating metrics connecting the near and far-field. We extend our previous work by computing the quasinormal modes of gravitational perturbations for static, spherically symmetric metrics that deviate from Schwarzschild spacetime. Even at the lowest order, this approach compares well with existing methods in both accuracy and applicability. Additionally, we show that the lowest-order interpolating metric reliably predicts light ring locations. It closely matches exact results, even when unsuitable for quasinormal frequency calculations.

		\medskip

	\end{abstract}
	
	\maketitle
	
	\section{Introduction}
	Three interferometric techniques are largely responsible for the identification of astrophysical black holes \cite{GEG:24}. Over the last decade, two of them have led to particularly spectacular breakthroughs. First, laser Michelson interferometry, through the efforts of the LIGO--Virgo--KAGRA collaboration, achieved the direct detection of gravitational waves  produced by the collisions of compact objects \cite{LIGO:23}. Second, in the electromagnetic spectrum, a black hole appears as a shadow — a dark spot against the image of a bright radiation source in the local sky. The Event Horizon Telescope, a global very long baseline interferometry array operating in the millimeter range, reconstructed event-horizon-scale images of this shadow for two of the most promising supermassive black hole candidates \cite{EHT:22}.

	Gravitational waves and black hole shadows are expected to serve as sensitive probes of the features predicted by various black hole models \cite{BCNS:19,CP:19}. As compact objects settle into equilibrium following violent astrophysical events, the late-time gravitational wave signal is dominated by quasinormal modes (QNMs) \cite{C:92,FN:98,N:99,BCS:09,KZ:11}. Both the ringdown to equilibrium and the shadow of an ultracompact object, along with the bright region bounding it, are closely tied to a special set of bounded null orbits known as photon or light rings (LRs) \cite{CMBWZ:09,BCS:09}.  The study of QNMs and LRs is crucial for probing the underlying spacetime geometry, offering insights into the physical properties of black holes and potential deviations from classical general relativity (GR), such as effects from quantum gravity or alternative theories of gravity \cite{BCNS:19,CP:19}.
	
	This is particularly significant as the question of the nature of astrophysical black holes is far from settled, and numerous models  generalizing the vacuum Schwarzschild and Kerr solutions of GR  have been proposed \cite{BCNS:19,CP:19}. Some are designed to address inherent deficiencies in the paradigmatic models, such as the presence of horizons and singularities. Others arise naturally from modified theories of gravity or the incorporation of quantum effects, or are otherwise designed to account for the presence of matter fields in the vicinity of the black hole. Among these possibilities, many models are indistinguishable in the asymptotic region, and only near the horizon do they differ significantly in their properties.
	
	While QNMs and LRs offer a potential means to distinguish between these models, their computation is typically model-specific and highly sensitive to underlying assumptions, such as the precise matter content and the nature of deviations from GR. Thus maximally model-independent descriptions that are able to encompass different classes of proposed models are particularly valuable \cite{CP:19}. These challenges were partly addressed in Refs.~\cite{RZ:14,KZ:22,KRZ:16}, which introduced a Padé-approximation-based framework for parameterizing black hole metrics, allowing for a systematic and broadly applicable study of QNMs and LRs.

	In recent work \cite{ST:24}, some of us explored an alternative way to construct approximating metrics which can be applied to a wide variety of relevant scenarios, based on a different Pad\'{e} interpolation scheme \cite{BGM:96,CPVWJ:08}. Motivated by a particular set of black hole models in semiclassical gravity whose metrics can only be determined in an asymptotic expansion near the horizon \cite{MMT:22}, two-point  Pad\'{e} functions were constructed which could effectively approximate the entire geometry by interpolating between the known near-horizon form of the metrics and some assumed asymptotic structure. This particular interpolating scheme, based on an M-fraction expansion, is advantageous over other parameterizations in that the coefficients describing the near and far regions are manifestly independent of each other, while also providing a complementary description in regions of the parameter space where metrics based on the C-fraction of Refs.~\cite{RZ:14,KZ:22} cannot be constructed.

	In this paper, we present the simplest class of Padé metrics useful for describing the observational characteristics of black holes. We show that these approximations are applicable not only for studying QNMs but also for analyzing LRs. They are designed to describe geometries that deviate slightly and continuously from the Schwarzschild solution without committing to a specific form or origin.
	First, we extend our previous work by computing gravitational perturbations of these Schwarzschild-like metrics. In Ref.~\cite{ST:24}, only scalar perturbations were considered, whereas gravitational perturbations are crucial for linking theoretical models with observational data. Here, we focus on Regge--Wheeler (axial or odd-parity) gravitational perturbations, which, in spherical symmetry, satisfy the same wave equation as scalar perturbations but with a different master potential.
	Second, we analyze the light rings of the resulting geometries, demonstrating how their location depends on the near-horizon geometry. We further illustrate the broader applicability of Padé metrics by reproducing the exact LR location of some well-known black hole models.

	This paper is organized as follows. In Sec.~\ref{p-sec}, we construct a variety of Padé metrics designed to approximate astrophysical black holes and discuss their relationship to other proposed parameterizations. In Sec.~\ref{qnms}, we derive the wave equation for axial gravitational perturbations of the Padé metric and compute the quasinormal frequencies and damping times for the $l=2$ fundamental mode, illustrating their dependence on near-horizon quantities. We benchmark our approach against established results in Sec.~\ref{qnm-bench}. In Sec.~\ref{light}, we determine the location of light rings for these metrics and compare them to those of the Schwarzschild black hole. We conclude with remarks on limitations and potential directions for future work.
	
	Throughout this article, we work in natural units such that $G=c=\hbar=1$. Where convenient, the coordinate distance from the Schwarzschild radius is denoted $x\defeq r-r_g$, as in \cite{MMT:22}, and the compactified radial coordinate is denoted $\bar x\defeq 1-r_g/r$. Note that the former quantity is denoted as $x$, and the latter as $\tilde{x}$ in Refs.~\cite{RZ:14,ST:24}.

	\section{Pad\'{e} metrics}\label{p-sec}
	In this Section, we provide a brief summary of the essential features of the Pad\'{e} approximation.  After discussing its general features as applied to spherically-symmetric metrics in Sec.~\ref{gen-pade}, we present the construction of Pad\'{e} approximations which describe metrics that deviate slightly from the Schwarzschild, Reissner--Nordström and Bardeen solutions (Secs.~\ref{schw-pade}--\ref{bar-pade}). We conclude in Sec.~\ref{rz-pade-KZ} with a comparison of M- and C-fraction expansions.
	
	\subsection{Pad\'{e} approximants}
	
	The multi-point Pad\'{e} approximation (constructed from finite-order elements of a Pad\'{e} expansion) is designed to approximate a function $f(z)$ which is specified by a finite power series expansion about a finite number of different points. For our purposes, we will be interested in the two-point  Pad\'{e} approximation. Suppose the expansion of the function $f(z)$ about $z=0$ and $z=\infty$ is known to some finite order and is given by
	\begin{align}
		\cF_0(z)=&\sum_{j=0}^{\infty}\, c_j z^j, \qquad\ \ \  c_j \in \mathbb{C}, \quad c_0 \neq 0\ ,\label{l1} \\
		\cF_{\infty}(z)=-&\sum_{j=1}^{\infty}\, c_{-j} z^{-j}, \quad c_{-j} \in \mathbb{C}, \quad c_{-1} \neq 0\ . \label{l2}
	\end{align}
	An approximation to the complete function can the be written as an M-fraction \cite{MM:76,CPVWJ:08}, a continued fraction given by
	\be\label{mfrac}
	M_\cF(z)= \frac{F_1}{1+G_1 z}\begin{array}{c}
		\\
		+\end{array}\ContFracOp_{k=2}^{\infty}\left(\frac{F_k z}{1+G_k z}\right)\ ,
	\ee
	where $F_k \in \mathbb{C} \backslash\{0\}$ and $G_k \in \mathbb{C}$, and the coefficients $\{F_k,G_k\}$ are determined from the $c_j$'s through the formulas given in Appendix \ref{app1}. The operator $\ContFracOp$ represents a continued fraction,
	\be
	\ContFracOp_{k=1}^{\infty}\left(\frac{a_k}{b_k}\right)=
	\cfrac{a_1}{b_1+\cfrac{a_2}{b_2+\cfrac{a_3}{b_3+\cdots}}} \ .
	\ee
	In practice the fraction is terminated at the $n$-th order, with $a_m=b_m=0$ for $m>n$. An equivalent description in terms of a rational function is given by
	\be
	\cF_n(z)=\frac{P^{[n-1]}(z)} {Q^{[n]}(z)} \ ,  \label{fn}
	\ee
	where $P^{[n-1]}(z)$ and $Q^{[n]}(z)$ are polynomials of order $n-1$ and $n$, respectively. The expansions of $\cF_n$  at $z=0$ and $z=\infty$ coincide with those of Eqs.~\eqref{l1} and \eqref{l2} up to terms of order $K_0$ and $K_\infty$, where $K\in \mathbb{Z}$ and   $K_0+K_\infty\leqslant 2n-1$. The orders $K_0$ and $K_\infty$ depend on the employed reconstruction algorithm.
	
	This two-point Padé technique has proven valuable in the study of various systems where dual asymptotic expansions exist, including certain supersymmetric Yang-Mills theories \cite{BT:13}, studies of critical phenomena \cite{G:20}, quantum mechanical stationary state problems \cite{LW:02}, and other systems exhibiting strong/weak coupling duality. Here, we will employ such Pad\'{e} approximations to describe the metric outside of the Schwarzschild radius $r_g$ of the presumed compact object. Clearly, if $Q^{[n]}(r)$ has one or more zeros at $r>r_g$, then the resulting interpolating function will be unsuitable for the computation of QNMs, and a different scheme is required \cite{ST:24}. We describe this issue in detail below.

	\subsection{Spherically-symmetric metrics: general considerations}\label{gen-pade}
	We now present the construction of the two-point  Pad\'{e} approximation for a number of static spherically symmetric black hole metrics. As we are primarily interested in presenting a new interpolation scheme and its application to the most common black hole models, we  consider   restricted  (static) line elements of the form
	\be
	ds^2=-f(r)dt^2+\dfrac{dr^2}{f(r)}+r^2d\Omega_2\ , \label{sssm}
	\ee
	where $d\Omega_2$ is the metric on the unit 2-sphere. While the most general static spherically symmetric metric in Schwarzschild coordinates has $g_{00}=-e^{2h(r)}f(r)$, for all metrics of Secs.~\ref{schw-pade}--\ref{bar-pade} we take $h\equiv 0$. In the static case the trapped region (with $f(r)<0$), if it exists, is bounded by the event horizon at some areal radius $r=r_g$. In all spherically-symmetric foliations the function $f(t,r)$ is coordinate-independent,
	\be
	f(t,r)\defeq 1-\frac{2M(t,r)}{r}\equiv 1-\frac{C(t,r)}{r} \defeq \pad_\mu\, r\pad^\mu r \ ,
	\ee
	where $M(t,r)=C(t,r)/2$ is the invariant Misner--Sharp--Hernandez (MSH) mass \cite{vF:15}, which in our case is function of only the radial coordinate.
	
	We are interested in functions which interpolate between the asymptotic region $r\sim\infty$ and the near horizon region $r\sim r_g$, where $r_g$ is defined as the outermost real root of $f(r)=0$ and represents the location of the event horizon.  The need for such a construction appears naturally when attempting to constrain the general form of a black hole metric in a semiclassical setting. Appendix~\ref{appSc} provides some details on the self-consistent semiclassical analysis of such metrics and their relevant properties.
	
	The equation $f(r)=0$ is assumed to have at least one real root at a finite areal radius $r=r_\sg$. The largest such root (on a non-cosmological scale) is the radial coordinate of the event horizon. The expansion of $f(r)$ near $r=r_g$ (denoted $f_{r_g}$) \cite{MMT:22,ST:24} is then
	\be
	f_{r_g}(r) =\sum_{k\geqslant 1}\frac{\alpha_k}{r_g^k}x^k\ , \label{exp-close}
	\ee
	where $x\defeq r-r_g$. We also assume that the spacetime is asymptotically flat and thus its far field expansion, which we denote $f_{\infty}$, is of the form
	\be
	f_{\infty}(r)=1-\sum_{k\geqslant 1}\frac{\beta_k}{\rho^k} \ , \label{exp-far}
	\ee
	with $\rho\defeq r/r_g$, and $\beta_1 r_g\equiv 2M=\lim_{r\to\infty} C(r)$, where $M$ is the  Arnowitt--Deser--Misner (ADM) mass. It is also to useful to introduce
	\be
	\epsilon\defeq\frac{2M}{r_g}-1 .
	\ee
	For example, for the Schwarzschild metric the dimensionless expansion coefficients are equal to $\alpha_k^\text{Schw}= (-1)^{k+1}$,  $\beta_1^\text{Schw}=1$, and $\beta_{k\geqslant 0}^\text{Schw}=0$, where additionally $\epsilon=0$. Neutron star compactness provides a loose lower bound of  $\epsilon \gtrsim -0.4$ \cite{KZ:22}.
	
	We construct  Pad\'{e} approximants of
	\be
	\cF(x)\defeq 1-f\big(r_g+x)=\frac{C(r_g+x)}{r_g+x} \ ,
	\ee
	whose expansion in $x$ has the form of Eqs.~\eqref{l1} and \eqref{l2} around $x=0$ and $x=\infty$, respectively.
	We label the order of the highest retained terms around these points as  $K_g$ and $K_\infty$, respectively. The order $n$ approximant of Eq.~\eqref{mfrac} has $2n$ independent parameters $(F_1,\ldots G_n)$. Our choice of $\cF$ automatically accommodates asymptotic flatness.  By definition of the horizon, $f(r_g)=0$, i.e., $\cF(0)=1$ and thus $F_1=1$. The remaining  $2n-1$ conditions are imposed by up to  $K_g+K_\infty$ coefficients $\alpha_i$ and $\beta_i$. As a result, the lowest order approximant that can capture  two terms of the near-horizon expansion and the leading large-distance behavior of $f$ corresponds to $n=3$,
	\be
	f_{3}(r)=1- \cfrac{1}{1+G_1x+\cfrac{F_2x}{1+G_2x+\cfrac{F_3x}{1+G_3x}}} \ . \label{f3cont}
	\ee
	It can also be represented as
	\be\label{f3schw}
	f_{3}(r)=\frac{A_3r^3 +A_2r^2 +A_1 r +A_0}{A_3 r^3+B_2 r^2 +B_1 r +B_0}\ ,
	\ee
	and is sufficiently versatile to incorporate many important features of solutions both without a global $U(1)$ charge as well as metrics of the Reissner--Nordström and Bardeen types.
	
	\subsection{{The minimal $n=3$ approximant}}\label{schw-pade}
	
	We begin with a presentation of the simplest ($n=3$) interpolating metric, where coincidence in the near and far-field expansions is enforced only up to terms of order $K_g=2$ and $K_\infty=1$, such that
	\begin{align}\label{schw-exp}
		f_{r_g}(r)&=\dfrac{\alpha_1}{r_g}x+\dfrac{\alpha_2}{r_g^2}x^2+\mathcal{O}(x^3)\ ,\\
		f_{\infty}(r)&=1- \dfrac{\beta_1}{\rho}+ {\mathcal{O}\left(\rho^{-2}\right)}\ ,\label{schw-exp2}
	\end{align}
	with higher-order coefficients left unconstrained.
	As an example and a benchmark we use the Schwarzschild metric
	\be
	f_{\text{Schw}}(r)=1-\dfrac{r_g}{r}\ ,
	\ee
	but as no feature of the far field behavior of the metric function beyond its decay as $1/r$ is enforced, it may be also used as a crude description of  other metrics.

	Using Eqs.~\eqref{mfrac} and \eqref{hankel1}-\eqref{hankel2}, we can obtain the $n=3$   Pad\'{e} interpolation of the expansions \eqref{schw-exp}-\eqref{schw-exp2}  where the parameters defining the interpolating metric are functions of  $\{r_g,\alpha_1,\alpha_2,\beta_1\}$ only, providing an alternative to the constructions of \cite{ST:24}. The coefficients $\{A_0,\dots, B_2\}$ of the polynomials in Eq.~\eqref{f3schw}, as well as the truncated continued fraction expansion are given in
	Appendix~\ref{a-detail}. Here we quote the representation of the $n=3$ interpolating metric as a rationalized fraction,
	\begin{widetext}
		\be
		f_3(r)=1-\frac{\beta_1r_g\big((1-2\alpha_1\beta_1-\alpha_2\beta_1^2) x^2 +\beta_1r_g(1-\alpha_1\beta_1 )x+\beta_1^2r_g^2\big)}
		{(1-2\alpha_1\beta_1-\alpha_2\beta_1^2)x^3+\beta_1r_g(1-\alpha_1\beta_1)x^2+\beta_1^2r_g^2x+\beta_1^3r_g^3}\ . \label{fS3}
		\ee
	\end{widetext}
	
	This approximant describes both the Schwarzschild geometry in the appropriate limit as well as small deviations from it as encoded by the near-horizon expansion coefficients $\{\alpha_1,\alpha_2\}$, where the ADM mass of the black hole is $2M=\beta_1 r_g$.   The interpolating function presented here is of a lower order than the $n=4$ function considered in our previous work \cite{ST:24} which, while also matching the near and far-field expansions \eqref{schw-exp}-\eqref{schw-exp2} to the same order as $f_3$, was erroneously believed to be the simplest possible interpolating function that   reproduces $f_{\text{Schw}}(r)$ where its parameters take the values of the Schwarzschild metric. This is because it may occur (as it does for the $n=3$ approximant) that while the continued fraction form of the approximant contains poles which superficially prevent one from obtaining a smooth limit to the Schwarzschild metric, upon rationalizing the continued fraction and taking the correct order of limits (which do not commute in general) one can recover a smooth Schwarzschild limit even when $n=3$.

	The coefficients $\alpha_1$ and $\alpha_2$ can be related to the energy conditions as described in Appendix~\ref{appSc}. This metric will be used in Sec.~\ref{qnms} and Sec.~\ref{light} to study the behaviour of gravitational perturbations and light rings for various black hole geometries.
	
	\begin{figure}[ht]
		\centering
		\includegraphics[width=0.42\textwidth]{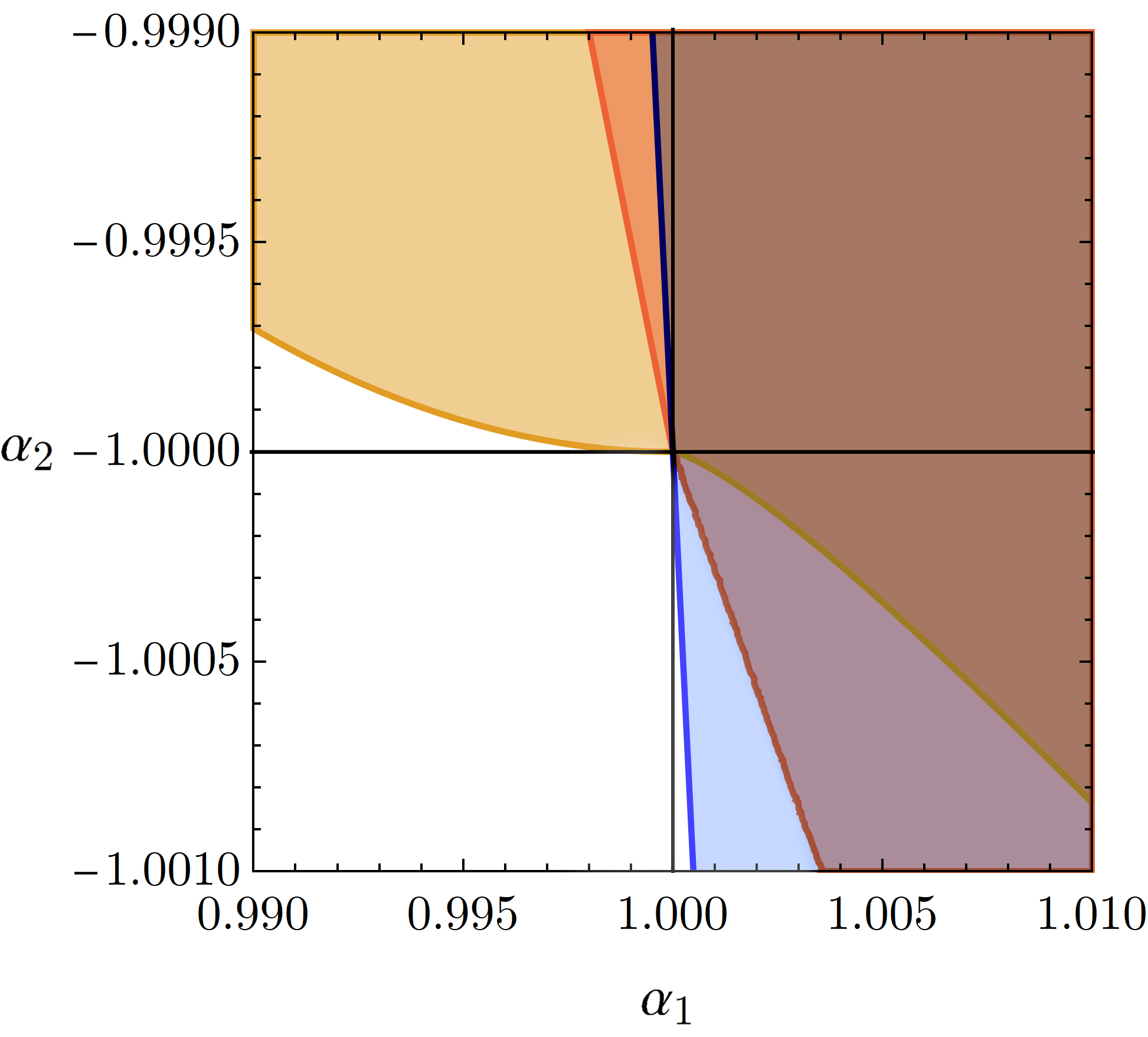}
		\caption{\small{The parameter space $(\alpha_1,\alpha_2)$ covered by the two-point
				Pad\'{e} approximations $f_{\text{S3}}$ (blue) and the approximations with $n=4$ (orange) and $n=5$ (yellow) of Ref.~\cite{ST:24}. In all expansions we set $\beta_1=1$. For the expansions $n=4$ and $n=5$ we take the Schwarzschild values of the higher order coefficients, $\alpha_3=1, \alpha_4=-1$. Poles in the respective function $f$  develop on the interval $r\in[r_g,\infty)$ for coefficients lying in the
				shaded region.}}\label{f-S3}
	\end{figure}

	We do not assume any particular theory or matter context as the source for the metric at this stage. Subleading terms in either of the asymptotic expansions are determined entirely by the set $\{\alpha_1,\alpha_2,\beta_1\}$, which can be taken to vary independently of each other while preserving agreement with the specified expansions up to the given order.

	\subsection{Reissner--Nordström-like metrics}\label{rn-pade}
	
	We now describe how the two-point  Pad\'{e} interpolation can be used to construct a four-parameter family of metrics describing deviations from the Reissner--Nordström  metric, which itself describes a spherically symmetric black hole spacetime with ADM mass $M$ and a single $U(1)$ gauge charge. The Reissner--Nordström metric in Schwarzschild coordinates is determined by
	\be
	f(r)=1-\dfrac{2M}{r}+\dfrac{q^2}{r^2}\ ,
	\ee
	which can be written in terms of the horizon radius $r_g$ as
	\be\label{rn}
	f(r)=1-\frac{r_g^2+q^2}{r\, r_g}+\frac{q^2}{r^2}\ .
	\ee
	We consider the expansions of Eqs.~\eqref{exp-close} and \eqref{exp-far} with $K_g=K_\infty=2$. The Reissner--Nordström metric itself is represented by
	\be\label{rnlimit}
	\alpha_1=1-\frac{q^2}{r_g^2}\ ,\qquad \alpha_2=\frac{2 q^2 }{r_g^2}-1\ ,
	\ee
	and
	\be
	\beta_1=1+\frac{q^2}{r_g^2}\ ,\qquad \beta_2=- \dfrac{q^2}{r_g^2}\ .
	\ee
	where the additional subleading term of order $r^{-2}$ must be included to account for the far-field effect of charge. Without this additional parameter, the interpolating function cannot reach the Reissner--Nordström solution as a limit.
	
	\begin{figure}[ht]
		\centering
		\includegraphics[width=0.41\textwidth]{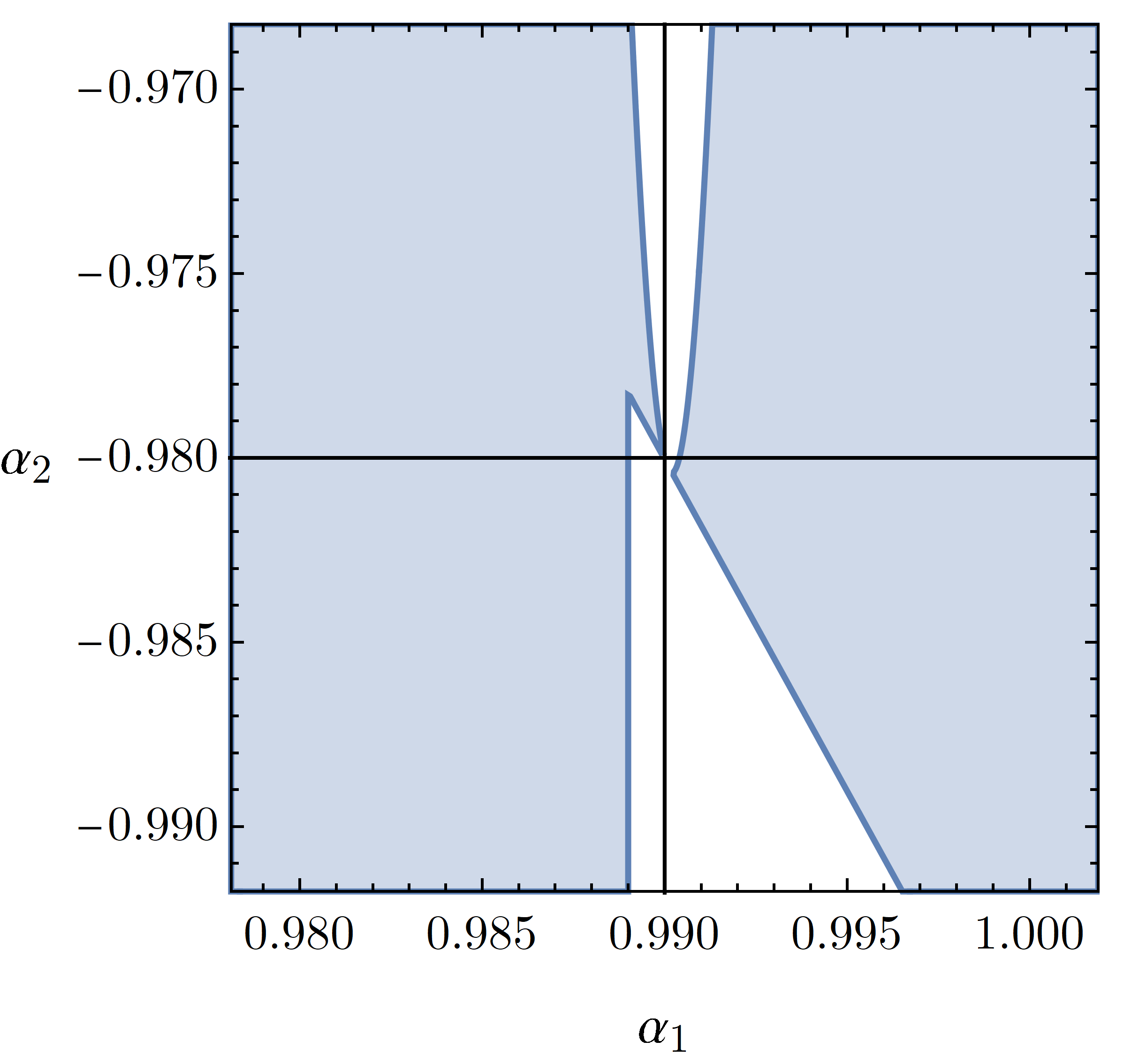}
		\caption{\small{The parameter space $(\alpha_1,\alpha_2)$ covered by the two-point Pad\'{e} approximation $f_{\text{RN3}}$ for $r_g=1$ and $q=0.1$. In this example, $\beta_1$ and $\beta_2$ are set to their Reissner--Nordström values. Poles on the interval $r\in[r_g,\infty)$  develop for coefficients lying in the shaded region.}}\label{f-RN3}
	\end{figure}
	
	The lowest order M-fraction approximation for which all four coefficients $\{\alpha_1,\alpha_2,\beta_1,\beta_2\}$ enter non-trivially is again $n=3$. The explicit form of $f_{\mathrm{RN}3}$ is given in the Appendix~\ref{a-detail}. Its domain of validity is shown in Fig.~\ref{f-RN3}.

	\subsection{Bardeen-like metrics}\label{bar-pade}
	
	\begin{figure}[ht]
		\centering
		\includegraphics[width=0.39\textwidth]{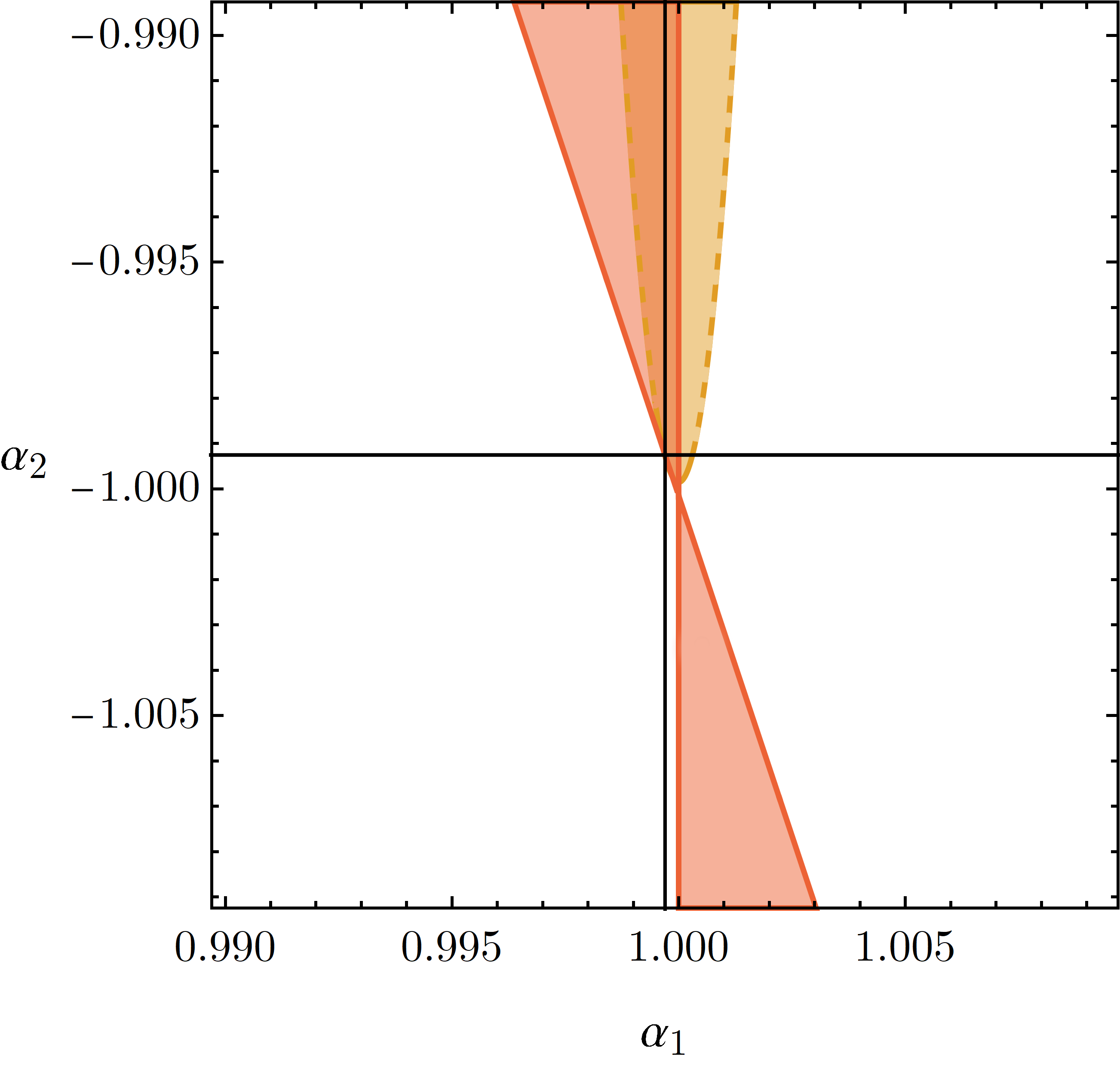}
		\caption{\small{Domains of validity for the lowest order M-fraction (yellow) and RZ (red) approximations of the Bardeen metric. Poles are present for the respective metrics in the shaded region. $\beta_1$ is fixed to its Bardeen value.}}\label{compare}
	\end{figure}
	
	The Bardeen metric is one of the oldest examples of a putative regular black hole --- a trapped domain of spacetime that does not contain singularity. Its metric is determined by
	\be
	f(r)=1-\dfrac{2Mr^2}{(q^2+r^2)^{3/2}}\ ,
	\ee
	{where, if the metric is sourced by nonlinear electrodynamics, $q$ is related to the magnetic charge $Q_m$ of the $U(1)$ gauge field given by $Q_{m}=(M q/2)^{1/2}$ \cite{MS:24}.}

	Rewriting it in terms of the Schwarzschild radius and the charge gives
	\be\label{bardeen}
	f(r)=1-\frac{r^2 \left(q^2+r_g^2\right)^{3/2}}{r_g^2 \left(q^2+r^2\right)^{3/2}}\ .
	\ee
	Hence the leading terms in  near and far field expansions of Eq.~\eqref{bardeen} are
	\begin{align}
		&\alpha_1=\dfrac{r_g^2-2q^2}{r_g^2+q^2}\ ,\qquad \alpha_2=\dfrac{11q^2r_g^2-2q^4-2r_g^4}{2(r_g^2+q^2)^2}\ ,\label{bdlimit1}\\
		& \beta_1=\dfrac{(r_g^2+q^2)^{3/2}}{r_g^3 }\ , \qquad \beta_3=-\dfrac{3q^2(r_g^2+q^2)}{2r_g^4 }\ ,\label{bdlimit2}
	\end{align}
	while $\beta_2=0$, i. e., the term of the order $1/r^2$ is absent in the far-field expansion \cite{MS:24}.
	
	To capture its characteristic behaviour we have to consider the expansions with $K_g=2$ and $K_\infty=3$. These still can be accommodated by an $n=3$ Pad\'{e} approximation. For a given set of $\{\alpha_1,\alpha_2,\beta_1,\beta_2,\beta_3\}$ the explicit form of the $n=3$ interpolating function $f_{\text{Bd3}}$ is given by Eq.~\eqref{f3cont} with the coefficients
	\begin{subequations}
		\begin{align}
			& F_2= \frac{\beta_1\alpha_1-1}{\beta_1 r_g} \ , \\
			& F_3= \frac{\beta_1 \left(\alpha_1^2 \beta_2-2 \alpha_1 \beta_1+\alpha_2 \left(\beta_2-\beta_1^2\right)+1\right)}{r_g (\alpha_1 \beta_1-1) \left(\beta_1^2-\beta_2\right)} \ , \\
			& G_1= (\beta_1 r_g)^{-1} \ , \\
			& G_2= \frac{\beta_1(1-\alpha_1\beta_1)}{r_g(\beta_1^2-\beta_2)} \ , \\
			&G_3= \frac{\left(\beta_2-\beta_1^2\right) \left(\alpha_1^2 \beta_2-2 \alpha_1 \beta_1+\alpha_2 \left(\beta_2-\beta_1^2\right)+1\right)}{r_g (\alpha_1 \beta_1-1) \left(\alpha_1 \beta_1 \beta_3+\alpha_1 \beta_2^2+\beta_1^3-2 \beta_1 \beta_2-\beta_3\right)} 
		\end{align}
	\end{subequations}
	The the Schwarzschild values of the coefficients $\alpha_1,2$ and $\beta_{1,2,3}$ the function $f_{\text{Bd3}}$ becomes the exact Schwarzschild metric function $f_{\text{Schw}}(r)$.
	
	It can be transformed into the rational fraction form of Eq.~\eqref{f3schw}. The pole structure of the approximant is determined by the zeroes  of the polynomial denominator
	\begin{widetext}
		\begin{align}
			Q_{\text{Bd3}}(r)&=r^3 \left(\alpha_1^2 \beta_2+2 \alpha_1 \beta_1+\alpha_2 \left(\beta_1^2+\beta_2\right)-1\right)-r^2 r_g \left(\alpha_1^2 (3 \beta_2+\beta_3)-\beta_1^2 (\alpha_1-3 \alpha_2)+\beta_1 (6 \alpha_1+\alpha_2 \beta_2+1)\right.\\
			&\left.\quad+\alpha_1 \beta_2+3 \alpha_2 \beta_2+\alpha_2 \beta_3-3\right)-r r_g^2 \left(\beta_3 \left(-2 \alpha_1^2+\alpha_1-2 \alpha_2\right)+\beta_1^2 (2 \alpha_1-3 \alpha_2+1)+\beta_1 (\alpha_1 (\beta_2-6)\right.\nonumber\\
			&\left.\quad+\alpha_2 (\beta_3-2 \beta_2)-2)-\beta_2 (\alpha_1 (3 \alpha_1+2)+\alpha_2 (\beta_2+3))+\beta_2+3\right)-r_g^3 \left(\beta_2 \left(\alpha_1^2+\beta_2 (\alpha_1+\alpha_2)+\alpha_1+\alpha_2-1\right)\right.\nonumber\\
			&\left.\quad+\beta_1^2 (-\alpha_1+\alpha_2-1)+\beta_1 (-\alpha_1 (\beta_2+\beta_3-2)+(\alpha_2+2) \beta_2-\alpha_2 \beta_3+1)+\beta_3 ((\alpha_1-1) \alpha_1+\alpha_2)+\beta_1^3+\beta_3-1\right)\nonumber
		\end{align}
	\end{widetext}
	Poles in $f_{\text{Bd3}}(r)$ are identified by any real roots of $Q_{\text{Bd3}}(r)$ that occur on the interval $r\in[r_g,\infty)$. These are absent for the Bardeen values of the coefficients given by Eqs.~\eqref{bdlimit1} and \eqref{bdlimit2}, but can appear when these coefficients  take  different values. Fig.~\ref{compare} shows the pole-free domain for various values of the parameter $q$.

	\subsection{Comparison to RZ Expansion}\label{rz-pade-KZ}
	
	The Rezzolla--Zhidenko (RZ) expansion \cite{RZ:14} employs the C-fraction Pad\'{e} approximants using a compact radial coordinate $\bar x\in[0,1)$ defined as
	\be
	\bar{x}\defeq 1-\dfrac{r_g}{r}\equiv 1-\frac{1}{\rho}\ .
	\ee
	It is applied to a general spherically-symmetric line element which can be written
	\be
	ds^2=-N^2(r) d t^2+\frac{B^2(r)}{N^2(r)} d r^2+r^2 d \Omega^2\ ,
	\ee
	so that the restricted class of metrics that we consider here corresponds to the choice $B\equiv 1$, i.e., $N^2(r)\to f(r)$.

	The metric function $N(r)$ is represented as
	\be
	N^2\eqdef\bar{x}A(\bar{x})\ ,\label{rzm0}
	\ee
	where
	\begin{align} \label{rzm}
		A(\bar{x})&= 1-\epsilon(1-\bar{x})+(a_0-\epsilon)(1-\bar{x})^2 \nonumber \\
		&+\tilde{A}(\bar{x})(1-\bar{x})^3\ ,
	\end{align}
	the function $\tilde{A}(\bar{x})$ is defined in terms of a C-fraction expansion as
	\be\label{RZcfrac}
	\tilde{A}(\bar{x})\equiv \cfrac{a_1}{1+\cfrac{a_2\bar x}{1+\cfrac{a_3\bar x}{1+\cdots}}}\ ,
	\ee
	with $\epsilon=\beta_1-1$ and $a_0=-\beta_2$. For uncharged black holes in GR
	\be
	a_0\equiv 0\ ,
	\ee
	while in a modified theory of gravity it is given as a difference of post-Newtonian parameters.  {Overall,  the expansion is determined by $m+3$ parameters: $r_g$, $\epsilon$, $\{a_i\}_{i=0}^m$. The two conditions $N^2(0)=0$ and $N^2(1)=1$ are enforced automatically, and the $m$-th order RZ expansion can reproduce $K_g+K_\infty=m+2$ near and far-field orders of the expansion of the metric function $f(r)$.}

	We illustrate the relation between the coefficients of the near and far field expansions of the metric and the RZ expansion for a case of $K_0=K_\infty=3$. The near-field coefficients
	are expressed as
	\begin{align}
		&\alpha_1 =  1 + a_0 + a_1 - 2 \epsilon\ ,  \label{a11}\\
		& \alpha_2 = -(1 + 3 a_0 + a_1 (4 + a_2) - 5 \epsilon) \ , \\
		& \alpha_3 = 1 + 6 a_0 +    a_1 (10 + a_2^2 + a_2 (5 + a_3) - 9 \epsilon),
	\end{align}
	and the far field coefficients are
	\begin{align}
		&\beta_1 =1 +\epsilon  \ , \\
		&\beta_2 =  -a_0 \, \\
		&\beta_3=a_0-\epsilon - \frac{a_1 (1 + a_3 + a_4)}{    1 + a_2 + a_3 + a_4 + a_2 a_4} . \label{be3}
	\end{align}
	The equations can be easily inverted to obtain the RZ parameters.
	
	Despite also being based on a continued fraction, there are important differences between the C-fraction-based construction of \cite{RZ:14} and the M-fraction expansion presented above. Both continued fractions  are truncated at some finite order, and the constants $\{a_i\}$ or $(F_i,G_i)$ appear  both in the near and far-field expansion of the resulting approximation of the known metric.
	
	However, the C-fraction naturally accommodates only the expansion near $z=0$. From one perspective the RZ construction allows to easily fix the far-field expansion up to the order $1/r^2$, and the higher order terms are defined by the near-horizon expansion coefficients $\alpha_{k\geqslant 3}$. While similar situation may arise in the M-fraction constructions, the algorithm to generate the coefficients $(F_k,G_k)$ are designed to deal with an arbitrary and independent number of the near and far coefficients. In addition, in the RZ construction polynomials $P$ and $Q$ in the polynomial fraction representation of the approximates that can incorporate the terms up to $K_0+K_\infty$ orders are typically of a higher order than for the M-fraction.

	It is useful to examine  the pole structure of both approximations to determine their suitability for describing small deviations about their parent metrics. This pole structure will differ significantly between the two constructions, evident immediately from the fact that for the same metric the two approximations yield rationalizations with different polynomials $Q^{[n]}$ and $Q^{[m]}$.
	
	We begin with the RZ construction applied to the Bardeen metric, truncating the C-fraction with $a_3=0$.  {Note that this approximate can match only the coefficients $\alpha_1$, $\alpha_2$ and $\beta_1=1+\epsilon$, $\beta_2=0$}.  In  the case $m=2$
	\be\label{rzbc}
	\tilde{A}(\bar x)=\dfrac{a_1}{1+a_2 \bar x}\ ,
	\ee
	{ and after setting $a_0\equiv 0$}  {we obtain}  the rational function
	\begin{widetext}
		\be\label{Brz}
		f_\mathrm{RZ3}(r)=\dfrac{(1+a_2)r^4+(\epsilon r_g-a_2 \epsilon r_g -2 a_2 r_g-r_g)r^3+(a_2 r_g^2+a_2 \epsilon r_g^2)r^2+(a_1r_g^3+\epsilon r_g^3+a_2 \epsilon r_g^3)r-(a_1r_g^4+a_2\epsilon r_g^4)}{(1+a_2)r^4-a_2r_g r^3}
		\ee
	\end{widetext}
	It is apparent from \eqref{rzbc} that when $a_2\leqslant  {- 1}$ the approximation develops a pole somewhere on the interval $  x\in[r_g,\infty)$ which is outside of the horizon.
	
	{We will now show that the parameters of the Bardeen metric itself is inaccessible for this expansion.}
	The values of the parameters corresponding to the Bardeen metric are
	\be\label{eps}
	a_1= -\frac{2 q^2 \sqrt{q^2+r_g^2}}{r_g^3}-\frac{2 \sqrt{q^2+r_g^2}}{r_g}+\frac{3 r_g^2}{q^2+r_g^2}-5\ ,
	\ee
	and
	\be
	\epsilon= -\frac{r_g^2 \sqrt{q^2+r_g^2}+q^2 \sqrt{q^2+r_g^2}+r_g^3}{r_g^3}
	\ee
	\begin{widetext}
		\be
		a_2=-\frac{3 \left(20 q^8 r_g^2+54 q^6 r_g^4+2 q^4 r_g^5 \left(5 \sqrt{q^2+r_g^2}+38 r_g\right)+q^2 r_g^7 \left(18 \sqrt{q^2+r_g^2}+59 r_g\right)+8 r_g^9 \left(\sqrt{q^2+r_g^2}+2 r_g\right)+3 q^{10}\right)}{\left(q^2+r_g^2\right) \left(39 q^6 r_g^2+67 q^4 r_g^4+57 q^2 r_g^6+9 q^8+24 r_g^8\right)}.
		\ee
	\end{widetext}
	{However}, the expansion has a pole outside the horizon since $a_2\leqslant -1$ for all values of $q$ and $r_g$.
	This limitation  {cannot}  be overcome by continuing the fraction to $a_3$ which  {both} allows $a_2\leqslant-1$  and captures the $r^{-3}$ term of the far-field expansion. The pole avoidance is ensured by two conditions
	\be
	a_2+a_3>-1, \qquad a_3>-1.
	\ee
	However, even in this case, as $q\rightarrow 0$ (which corresponds to the Schwarzschild limit)  we have
	\be
	a_3+a_2=-2+\frac{29 q^2}{6 r_g^2}+\cO(q^4/r_0^4) \,
	\ee
	indicating presence of a pole. In Ref.~\cite{KOA:23} the fifth and the sixth order Pad\'{e} approximations are used.

	\section{Quasinormal modes}\label{qnms}
	After a brief review of the essential elements of the QNM theory in Sec.~\ref{qnm-intro} we describe in Sec.~\ref{eq-sec} how Pad\'{e} approximants are used in their calculations. Finally, in Sec.~\ref{qnm-bench} we present some of the results that are based on using the $n=3$ approximants and compare them with their counterparts in the RZ scheme as well as with the established results for the Schwarzschild, Reissner–Nordstr\"{o}m, and Bardeen metrics.

	\subsection{Setting up the problem}\label{qnm-intro}
	The study of QNMs is extensively covered in the literature. Excellent reviews can be found in \cite{C:92,BCS:09,N:99,KZ:11,C:92,FN:98}, and here we provide only a brief summary of the relevant equations.

	We are interested in gravitational perturbations on a stationary spherically-symmetric background $g_{\mu\nu}\equiv g^0_{\mu\nu}$. Assuming the dynamics of interest occur in the linearized regime, small perturbations about the background metric $g_{\mu\nu}^0$,
	\be\nonumber
	g_{\mu\nu}=g_{\mu\nu}^0+\delta g_{\mu\nu}\ ,
	\ee
	where $|\delta g_{\mu\nu}|\ll |g_{\mu\nu}^0|$ in a suitable gauge. Perturbations of the fields of spin $s=0,1$ are
	\be
	\Phi=\Phi^0+\delta\Phi\ ,
	\ee
	where indices are suppressed.
	
	Both the gravitational perturbations (treated as spin $s=2$ fields), and the matter fields can be decomposed as
	\be
	\Psi^\#(t,r,\theta,\phi)=\sum_{lmn}\dfrac{1}{r}\,\psi_{lmn}(r) Y^\#_{lm}(\theta,\phi)e^{-i\omega_{lmn} t}\ , \label{ansatz}
	\ee
	where $l\geqslant s$ is the angular momentum number, $m$ is the azimuthal number, $n$ is the overtone number that characterizes the complex part of the frequency \cite{BCS:09,KZ:11}. Parity of the solution is determined by $(-1)^{l+1}$. We suppressed the indices, but the mark $\#$ indicates the scalar, vector or tensor type of the perturbation with $Y^\#_{lm}$ being the corresponding spherical harmonics \cite{N:99}. For spherically-symmetric backgrounds that we considered here it is enough to consider only $m=0$ perturbations \cite{C:92}.
	
	The ansatz \eqref{ansatz}, and imposition of the Regge--Wheeler gauge reduce the radial perturbation equations to the Schrödinger-like form  \cite{EV:70,N:99,BCS:09,KZ:11}
	\be\label{kg1}
	\dfrac{\p^2\psi}{\p r_*^2}+(\omega^2-V)\psi=0\ ,
	\ee
	where the tortoise coordinate \cite{FN:98,BCS:09,HE:73} $r_*$ was introduced via
	\be
	dr_*=  \frac{dr}{f(r)}\ .
	\ee
	Only two potentials $V(r)$, corresponding to odd (axial) and even (polar) perturbations, need to be considered. These are the Regge--Wheeler and Zerilli potentials, respectively. Adapting the results of \cite{EV:70,SNS:03,BNV:13} for the metric of Eq.~\eqref{sssm} we have
	\be
	V_\mathrm{RW}^s=f(r)\left(\frac{l(l+1)}{r^3}-s(s-1)\frac{2M(r)}{r^3}+(1-s)\frac{f'(r)}{r}\right) \ ,
	\ee
	for the Regge--Wheeler potential. The Zerilli potential can be derived in a similar fashion (see e.g. \cite{C:99,K:10}) and is significantly more complicated. It is worth noting that $V_\mathrm{RW}^s$ is also valid not only for black holes, but for ultracompact objects that are modelled with anisotropic perfect fluids \cite{BNV:13}.
	
	Despite their different functional forms the two potentials are very close in their actual values at least for the low values of $l$ \cite{N:99}. Moreover, in spherical symmetry the more complicated Zerilli equation can be transformed to the Regge--Wheeler equation \cite{C:92}. For the Schwarzschild metric the Regge--Wheeler and Zerilli potentials are known to be isospectral, thus giving the same QNMs.  While the implied isospectrality does not hold in general, it is still useful (and simpler) to consider the Regge--Wheeler perturbations first, as we do here.  We will be interested in the frequency domain only.
	
	In what follows it is convenient to work instead with the radial coordinate $r$, for which the radial equation \eqref{kg1} is
	\be\label{radial1}
	f^2\,\dfrac{\p^2\psi}{\p r^2}+ff'\dfrac{\p\psi}{\p r}+(\omega^2-V)\psi=0\ .
	\ee
	In a black hole background, the relevant solutions describing QNMs are a subset of Eq.~\eqref{radial1} satisfying purely ingoing boundary conditions at the horizon,
	\be\label{bcin}
	\psi\sim e^{-i\omega r_*}\quad \text{as}\quad r_*\rightarrow -\infty\ ,
	\ee
	and purely outgoing boundary conditions at infinity,
	\be\label{bcout}
	\phi\sim e^{i\omega r_*}\quad \text{as}\quad r_*\rightarrow +\infty\ .
	\ee
	
	Enforcing these boundary conditions selects solutions that correspond to damped quasinormal oscillations arising from an initial perturbation (hence the name QNM). The frequency of these oscillations is independent of the initial amplitude. They are characterized by the physical frequency $f$ and damping time $\tau$ of each damped sinusoid present in the signal, and are related to the complex frequency $\omega$ through
	\be\nonumber
	\omega_{lmn}=2\pi f_{lmn}+\dfrac{i}{\tau_{lmn}}\ .
	\ee
	
	In the general setting of an axially symmetric background it is still true that for fixed $a= J/M$ the complex frequency scales as $\omega\sim M^{-1}$, so it is convenient to define a dimensionless frequency
	\be
	\Omega_{lmn}=M\omega_{lmn}\ .\nonumber
	\ee
	A theory-independent framework for characterizing deviations from the vacuum Kerr solution's predicted frequency and damping time has been developed in \cite{L:12,GVS:12,G:21}. This approach assumes that quasinormal frequencies depend not only on the ADM mass and angular momentum $(M,J)$ but also on additional dimensionless parameters or charges ${\Delta \hat{\omega}_{lm}}$. These extra parameters could arise from either matter/gauge field interactions or modifications to the Einstein--Hilbert action. Within this framework, the observed frequency and damping time are expressed as fractional deviations from their Kerr values.:
	\begin{align}\nonumber
		& f_{l m 0}=f_{l m 0}^{\mathrm{GR}}\left(1+\delta \hat{f}_{l m 0}\right), \\
		& \tau_{l m 0}=\tau_{l m 0}^{\mathrm{GR}}\left(1+\delta \hat{\tau}_{l m 0}\right) \ .\nonumber
	\end{align}
	As described in \cite{A:22}, a detailed study of high-SNR black hole binary coalescence signals allows one to constrain the fractional deviations from the Kerr value of the $(220)$ mode to be
	\be\label{constraint}
	\delta \hat{f}_{220}=0.02_{\,-0.03}^{\,+0.03} ,\quad \delta \hat{\tau}_{220}=0.13_{\,-0.11}^{\,+0.11}\ .
	\ee
	where the bounds encompass the 90\% confidence interval.  {We will be interested in relating deviations in quasinormal frequencies of a similar order of magnitude to changes in the underlying metric parameters near the horizon, now focussing on gravitational perturbations (whereas our previous work \cite{ST:24} examined scalar perturbations as a proof of concept).}

	The boundary conditions of Eqs.~\eqref{bcin} and \eqref{bcout} only require the asymptotic form of the metric. Hence its behaviour of the metric near the horizon that is given by Eq.~\eqref{exp-close} and in the asymptotic region are sufficient to specify an ansatz to the wave equation \eqref{radial1}. For all spherically-symmetric metrics the leading term in the far-field expansion $(r\rightarrow \infty$) is
	\be\label{far}
	f(r)=1-\frac{2M}{r}+\cO(r^{-2}),
	\ee
	where $M$ is the ADM mass. It does not distinguish between scenarios with various sub-leading far-field behaviour. However, to obtain the effective potential $V$ it is necessary to define the metric over the entire radial domain, hence the pole structure of the Pad\'{e} approximations becomes a central issue.

	\subsection{QNM with the Pad\'{e} approximants}\label{eq-sec}
	
	Equation \eqref{radial1} along with the boundary conditions of Eqs. \eqref{bcin}-\eqref{bcout} furnish a standard perturbation problem in spherical symmetry. To solve it, we construct an ansatz which has the correct behaviour at the boundaries. We describe the case for the metric \eqref{f3schw} but the same procedure can be used to construct the radial wave equation for all metrics considered. Near the horizon the tortoise coordinate takes the form
	\begin{align}
		r_*&=\int\!\dfrac{dr}{f(r)}\underset{r=r_g}{\approx}\frac{r_g\log (r-r_g)}{\alpha_1}-\dfrac{\alpha_2 (r-r_g)}{\alpha_1^2}\ ,
	\end{align}
	where only terms up to order $\mathcal{O}(\tilde x)$ are kept. This can be seen to reduce to the Schwarzschild form in the limit $\alpha_1\rightarrow 1$ and $\alpha_2\rightarrow -1$,
	\be
	r_*=r+r_g\log\left(\frac{r-r_g}{r_g}\right)\ .
	\ee
	Near the horizon we therefore seek purely ingoing solutions of the form
	\begin{align}
		\phi\sim e^{-i\omega r_*}&=\left(r-r_g\right)^{-\frac{i w r_g}{\alpha_1}}e^{-i w \chi}R(r)\ ,
	\end{align}
	where we have defined
	\be
	\chi\equiv\-\frac{\alpha_2 (r-r_g)}{\alpha_1^2}\ .
	\ee
	The boundary condition at infinity is
	\be
	\phi\sim e^{i\omega r_*}=e^{i\omega r} \left(\dfrac{r}{r_g}\right)^{i w \beta_1  r_g }R(r)\ ,
	\ee
	The following ansatz achieves these limiting forms while possessing the correct limit when the coefficients $\alpha_1$, $\alpha_2$, and $\beta_1$ take on their Schwarzschild values:
	\be
	\phi(r)= e^{-i\omega \chi}(r-r_g)^{-\tfrac{i\omega r_g}{\alpha_1}}e^{2i\omega r}\left(\frac{r}{r_g}\right)^{\!i\omega\beta_1 r_g}r^{\tfrac{i\omega r_g}{\alpha_1}}R(r)
	\ee
	Inserting the ansatz into \eqref{radial1} and using \eqref{f3schw} and \eqref{kg1} reduces the radial equation to the form
	\be\label{radial2}
	\gamma(r)R''(r)+\tau(r)R'(r)+\sigma(r)R(r)=0\ ,
	\ee
	where the functions $\gamma$, $\tau$ and $\sigma$ are given in Appendix \ref{appB}.
	
	Before determining the coefficients  we transform the radial domain to a compact interval by using $\bar x\defeq 1-r_g/r$.
	We further let $\tilde{R}(\bar x)=\bar x(1-\bar x)R(\bar x)$ so that the boundary conditions become $\tilde{R}(0)=\tilde{R}(1)=0$. The radial equation \eqref{radial2} becomes
	\begin{align}
		\gamma(\bar x)\tilde{R}''(\bar x)+\tau(\bar x)\tilde{R}'(\bar x)+\sigma(\bar x)\tilde{R}(\bar x)=0\ ,
	\end{align}
	where primes indicate derivatives with respect to the argument $\bar x$.
	
	{When $\alpha_1=1 $, $\alpha_2=-1$, and $\beta_1=1$, the radial equation reduces to
		\begin{align}
			&\left[\frac{ l^2+l-8 r_g^2 \omega ^2+8 i r_g \omega +3-2 i r_g \omega -1}{(x-1)^3 x}\right.\nonumber\\
			&\left.-\frac{ \left(l^2+l-3 (i-2 r_g \omega)^2\right)+x (i-2 r_g \omega)^2}{(x-1)^3 }\right]\tilde{R}(\bar x)\nonumber\\
			&+\frac{ \left(x^2 (-1-4 i r_g \omega )+x (2+8 i r_g \omega )-2 i r_g \omega -1\right)}{(x-1)^2 x}\tilde{R}'(\bar x)\nonumber\\
			&+\tilde{R}''(\bar x)=0\ ,
		\end{align}
		which is the known form of the radial equation for odd parity gravitational quasinormal perturbations on the Schwarzschild background.}

	\subsection{Selected numerical results and benchmarking}\label{qnm-bench}
	
	Here we present a sample of the QNM frequencies of scalar and gravitational perturbations that are obtained for several $n=3$ Pad\'{e} M-approximants that we described in Sec.~\ref{p-sec} using the matrix method of Refs.~\cite{LQ:16,LQ:17}. We report the QNM frequencies as dimensionless quantities in the units of $r_g^{-1}$ and present only one member of the complex-conjugated pair. This is in agrement with the convention of Ref.~\cite{C:92} and twice the convention of Ref.~\cite{FN:98}.

\begin{figure}[ht]
		\centering
		\ \ \,\includegraphics[width=0.44\textwidth]{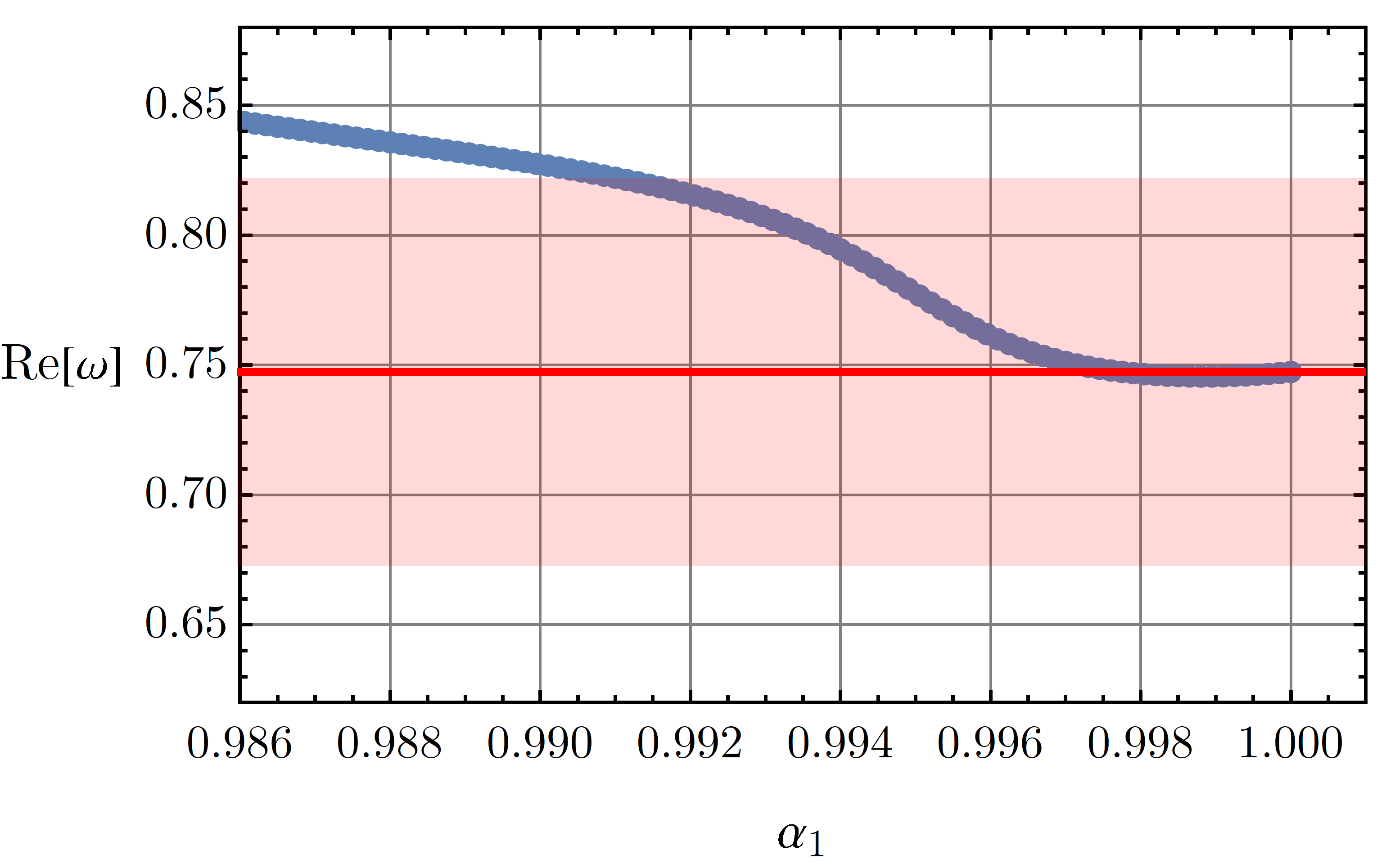}\\
		\includegraphics[width=0.45\textwidth]{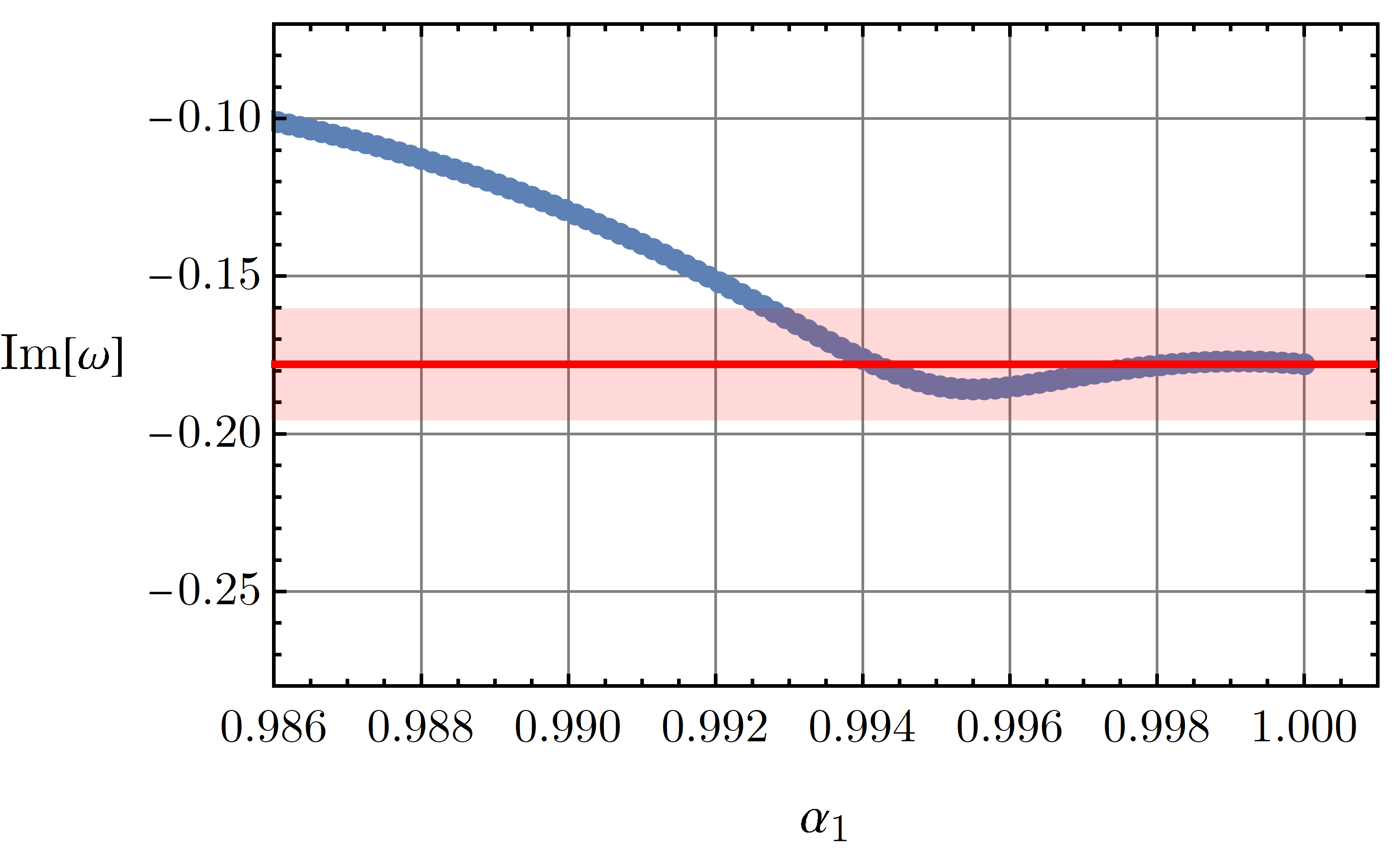}\\
		\includegraphics[width=0.45\textwidth]{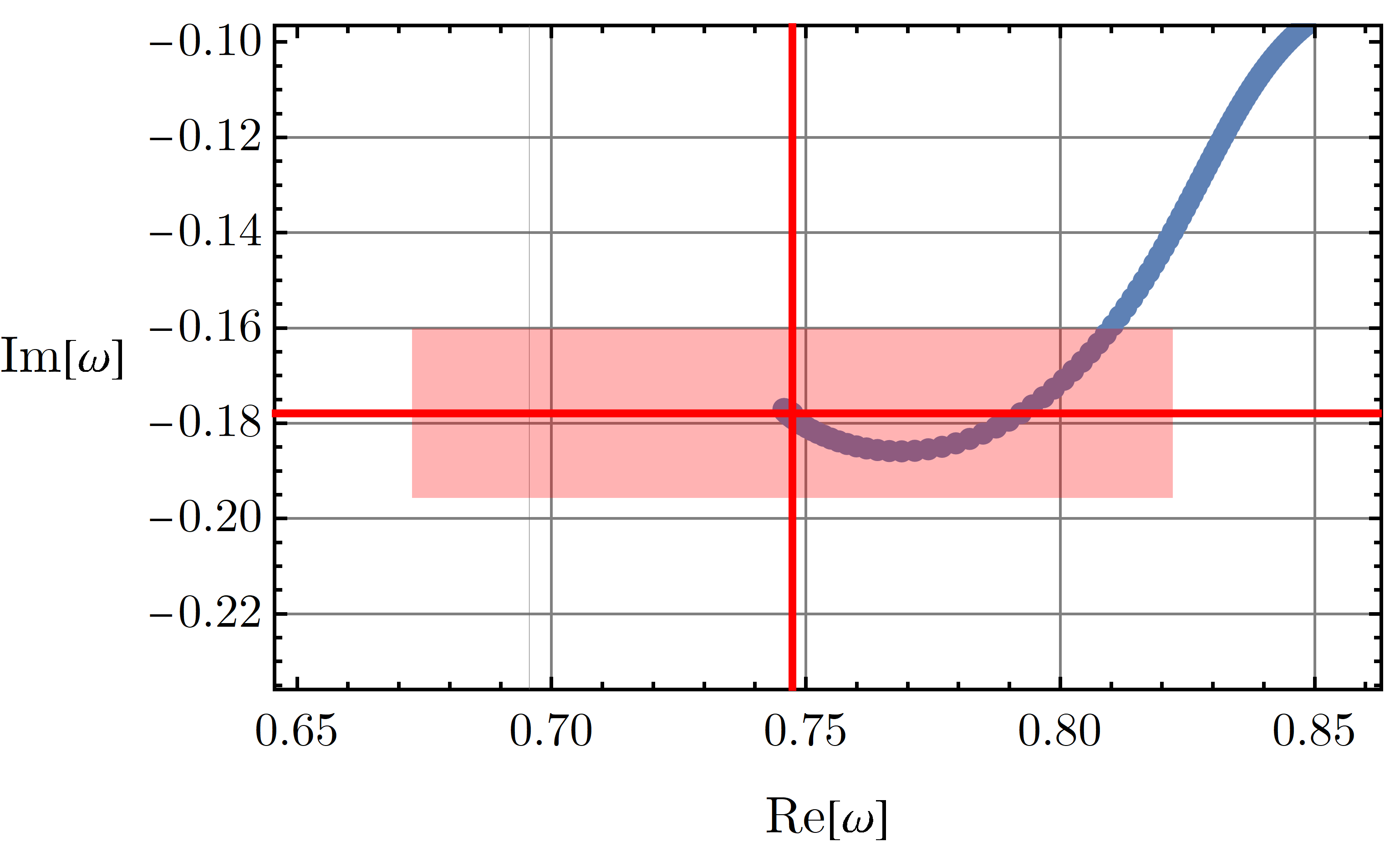}
		\caption{\small{Real, imaginary, and parametric plots of the fundamental $l=2$ mode of the $f_{3}$ metric as a function of $\alpha_1$. The step size is $\Delta\alpha_1=0.00015$. The shaded red region corresponds to a $10\%$ deviation window centered on the known Schwarzschild quasinormal frequency.}}
		\label{ga1}
	\end{figure}

\begin{figure}[ht]
		\centering
		\ \ \includegraphics[width=0.45\textwidth]{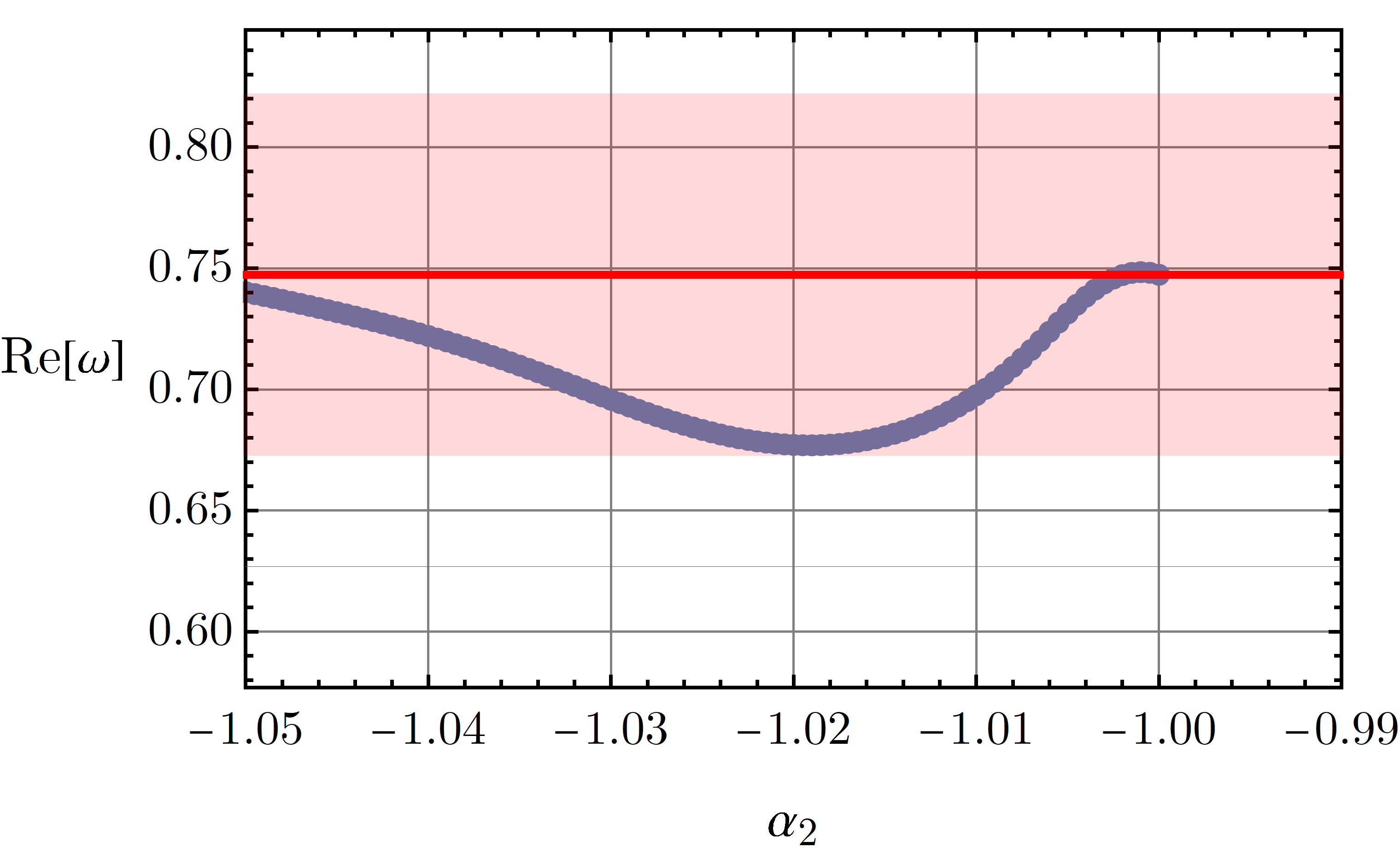}\\
		\includegraphics[width=0.46\textwidth]{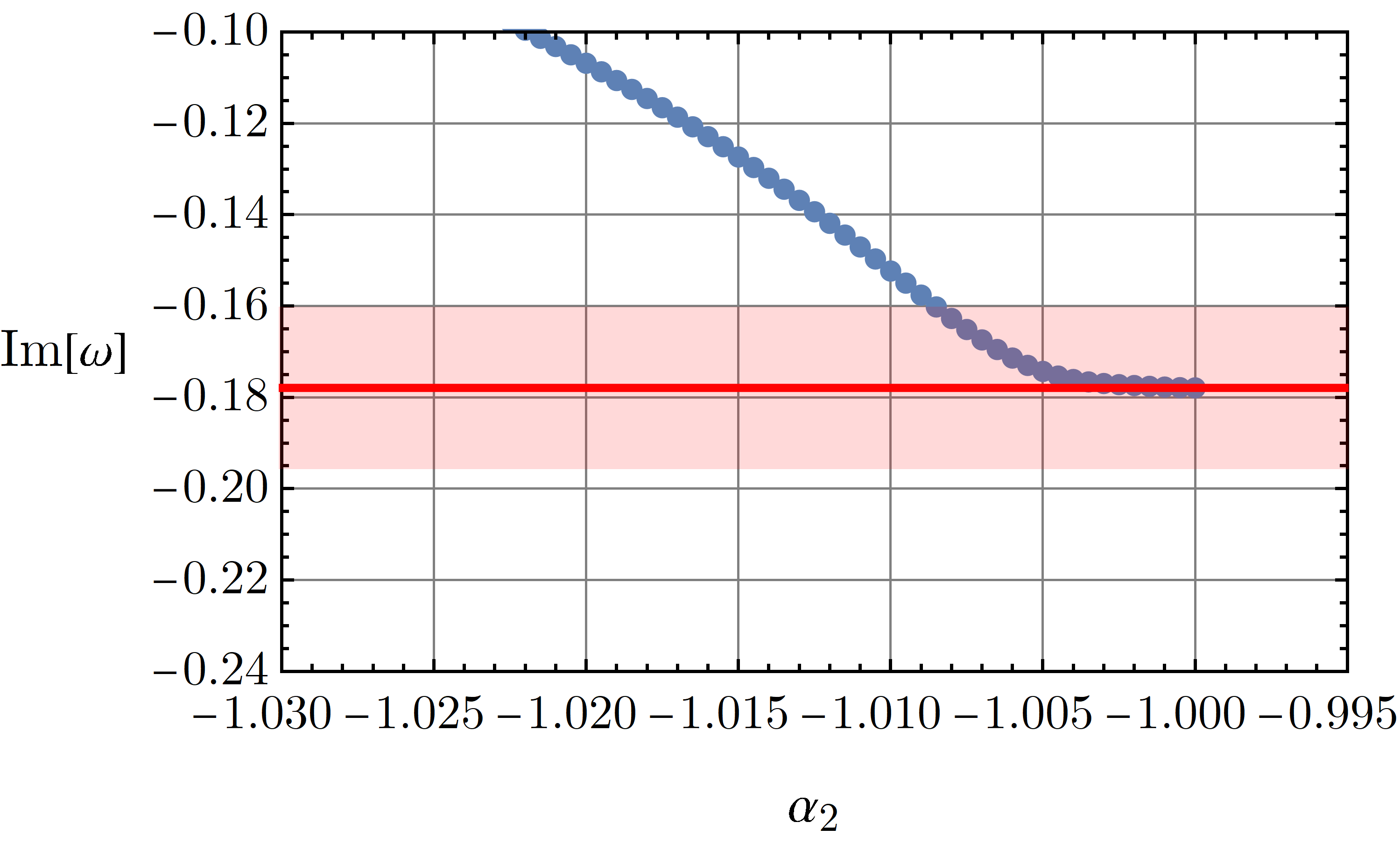}\\
		\!\!\includegraphics[width=0.456\textwidth]{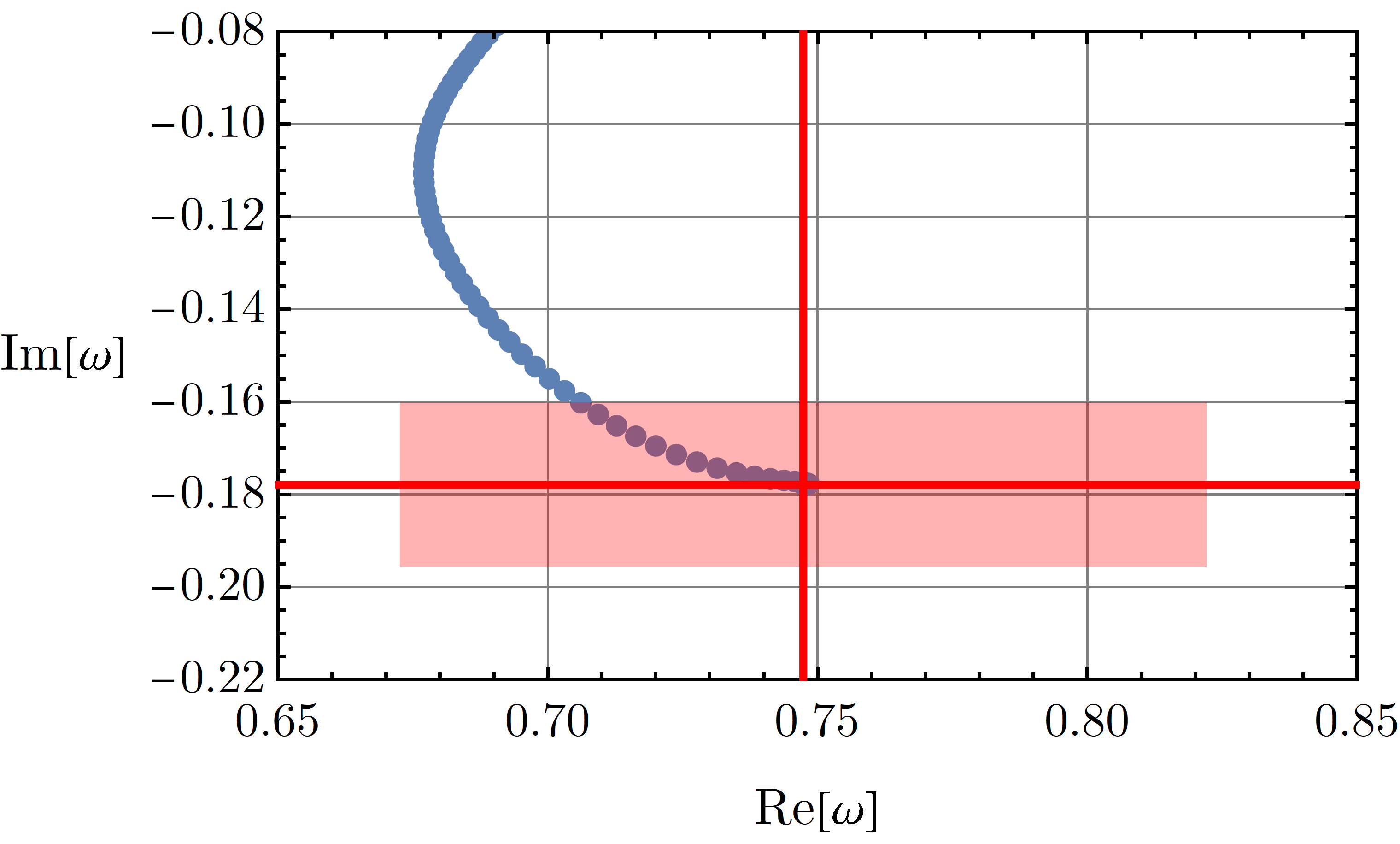}
		\caption{\small{Real, imaginary, and parametric plots of the fundamental $l=2$ mode of the $f_{3}$ metric function as a function of $\alpha_2$. The step size is $\Delta\alpha_2=0.0005$. The shaded red region corresponds to a $10\%$ deviation window centered on the known Schwarzschild quasinormal frequency.}}
		\label{ga2}
	\end{figure}

	We now report the behaviour of the $l=2$ fundamental mode $(200)$ of the gravitational perturbations for the metric $f_3$ of \eqref{f3schw}, which describes deviations from the Schwarzschild metric in the near-horizon regime as parameterized by small changes in the leading and subleading coefficients of  $\{\alpha_1,\alpha_2\}$. Here we keep $\beta_1\equiv 1$.   For the Schwarzschild metric  this QNM has frequency
	\be
	\omega_{200}=0.747344 - 0.177924 i
	\ee
	as determined by both pseudo-spectral and asymptotic iteration methods \cite{M:22}, and is indicated by the red lines in Fig.~\ref{ga1} and Fig.~\ref{ga2}. (This value provides one more significant figure relative to the standard results of Refs.~\cite{C:92,FN:98}).

	It is evident in both figures that the damping time is much more sensitive than the frequency to changes in both the leading ($\alpha_1$) and subleading ($\alpha_2$) terms in the near-horizon expansion of \eqref{f3schw}. This behaviour was also observed for the (100) mode in \cite{ST:24} for scalar perturbations, but not for (000) or (200). In fact, deviations in the subleading $\alpha_2$ term can likely only be reliably captured by the damping time of the mode, as the frequency remains within the specified 10\% window well past the point at which the small-$\alpha$ approximation becomes invalid.

	Next we compare the fundamental $l=0$ frequency for the QNM of gravitational perturbations for both the M-fraction expansion and RZ parameterization the Reissner--Nordström metric and small deviations from its leading parameters. First we note the frequencies for the Reissner--Nordström metric of Eq.~\eqref{rn} that are evaluated by Leaver \cite{L:90} are
	
	\begin{align}
		\omega^{(L)}_{200}&=-0.17792-0.74734i\quad(q=0.1), \\
		\omega^{(L)}_{200}&= -0.17880 -0.75687i \quad(q=0.2),\\
		\omega^{(L)}_{200}&= -0.17929 -0.78024i\quad(q=0.4),
	\end{align}
	
	for the three different values of the parameter $q$. For the metric $f_\mathrm{RN3}$  using a grid of $N=20$ we obtain an excellent agreement with
	
	\begin{align}
		\omega_{200}&=-0.17801-0.74772i\quad(q=0.1),\\
		\omega_{200}&= -0.17871 -0.75698i \quad(q=0.2),\\
		\omega_{200}&= -0.17945 -0.78061i\quad(q=0.4).
	\end{align}
	
	We also compare the RZ and M-fraction approximations in a region of overlapping validity, i.e. choosing a metric for which both approximations are valid. Using a grid with $N=30$ we find that
	
	\begin{align}
		\omega^{(RZ)}_{200}&=0.20851-0.52570i\\
		\omega^{(M)}_{200}&=0.20859-0.52562i\ .
	\end{align}
	
	Therefore, when both approximations are regular and equal up to second order in their respective near-horizon and asymptotic expansions, their respective quasinormal modes also agree to within one part in $10^{-4}$. Presumably, simultaneously taking higher orders of approximation for both schemes improves the agreement. Therefore, the schemes are comparable and achieve a similar level of precision when the relevant choice of parameters can be described by both.
	
	Finally, we compare our results for the Bardeen metrics with the recent calculations that were reported in Ref.~\cite{KOA:23}, which employed several high-order approximations. This work studied perturbations of $s=(0,\tfrac{1}{2}, 1)$ as opposed to gravitational perturbations of the Bardeen metric, using 6th-order WKB (for the exact metric and the 5th and the 6th order Pad\'{e} approximations) and 20th/24th-order Frobenius methods.  The  results are obtained by extracting the quasinormal frequency from a time-domain profile for the perturbation, in contrast to our frequency domain approach.

	Here we compare the fundamental frequency of the scalar perturbations for the  approximating metric $f_{\text{Bd3}}(r)$. We point out that the $l_0$ of Ref.~\cite{KOA:23} is labelled $q$ in our notation, and we consider the first few fundamental scalar modes given in  Table I. We use a discretization grid of $N=25$ points and modify the wave equation appropriately for the scalar case, obtaining the following values for the fundamental scalar quasinormal mode:
	\begin{align}
		\omega_{000}&=0.22094 - 0.20974i\qquad (q=l_0=0)\\
		\omega_{000}&=0.22079 - 0.20722i \qquad (q=l_0=0.07698)\\
		\omega_{000}&=0.21513 - 0.19492i \qquad (q=l_0=0.15396)
	\end{align}
	As the Bardeen metric Eq.~\eqref{bardeen} reduces to the Schwarzschild metric for $q=0$, here we quote the result for the fundamental scalar mode that is adapted from \cite{M:22},
	\be
	\omega_{000}^\mathrm{Schw}=0.220910 - 0.209792i.
	\ee
	Our results compare favourably to both the WKB approach and those obtained using a 24th order Frobenius method.  It is important to note that $f_{\text{Bd3}}(r)$, while numerically very close to the Bardeen metric in the limit where the coefficients take on their Bardeen values, is not exactly equal to it.  Nonetheless, the numerical difference between $f_{\text{Bd3}}(r)$ and the Bardeen metric is on the order of $10^{-9}$ and thus we are still able to compute the quasinormal modes with a good accuracy and a low computational cost. The metric does, however, reduce to the Schwarzschild metric exactly when $q=l_0=0$, and thus the frequency for this mode matches the expected value.

	\section{Light rings}\label{light}
	
	Beyond gravitational wave detection, electromagnetic signatures from the photon sphere—such as light rings and shadows—provide essential observational tools for investigating the true nature of astrophysical black hole candidates. Parameterizing deviations from standard black hole solutions \cite{CP:19, WCJ:17} enables comparisons across a continuum of models, while continued fraction expansions offer greater flexibility \cite{YZRKM:16}.

Additionally, Padé approximants explicitly reveal how light ring properties depend on parameters describing the near-horizon geometry. In this section, we use the simplest  $n=3$ metric function of Eq.~\eqref{fS3} and find the location of the LRs. In light of the results of Section~\ref{qnms} we work with the parameters $\alpha_i$ and $\beta_i$, which are assumed to be close to their Schwarzschild values. We show that, as it is intuitively expected, the LRs are insensitive to far field deviations in the metric.  While it seems reasonable that  the LR location would be influenced also by the terms $\alpha_{i\geqslant 3}$, a comparison with the exact solutions for the Reissner--Nordström and Bardeen metrics show that even a simpler approximation $f_\mathrm{S3}$ of Eq.~\eqref{fS3}   provides excellent estimates.

	Determining the location of LRs involves finding the circular null geodesics of photons \cite{C:92} in the given background geometry. Starting from the geodesic equation
	\be
	\tensor{\sg}{_\mu_\nu}\dot{x}^{\mu}\dot{x}^{\nu} = 0\ ,
	\label{eq:nc}
	\ee
	one obtains (in the static spherically-symmetric case) the equation
	\begin{align}
		\dot{r}^2 + V(r) = 0\ ,
		\label{eq:V}
	\end{align}
	where the derivative is taken over the affine parameter and the effective potential $V(r)$ depends on the conserved energy $E$ and the angular momentum $L$ as
	\be
	V(r) \defeq \frac{E^2}{\tensor{\sg}{_t_t} \tensor{\sg}{_r_r}} + \frac{L^2}{\tensor{\sg}{_r_r} \tensor{\sg}{_\phi_\phi}}\ .
	\ee
	
	The location of the LRs is determined by the conditions $\dot{r}=0$ and $\ddot{r}=0$, which implies that $V(r)=0$ and $V'(r)=0$, respectively \cite{cbh:17,ch:20}. Solving these equations for $r$ gives the radius of the LRs and the impact parameter $b \coloneqq L/E$.
	
	An alternative method, which is more convenient for our purposes, is to identify the LR locations by determining the critical points of the function (see Ref.~\cite{cbh:17} for a detailed derivation)
	\begin{align}
		H = \frac{-\tensor{\sg}{_t_\phi}\pm \sqrt{\tensor{\sg}{_t_\phi}-\tensor{\sg}{_t_t}\tensor{\sg}{_\phi_\phi}}}{\tensor{\sg}{_\phi_\phi}} \ , \label{eq:H-gen}
	\end{align}		
	i. e., finding the roots of the equation
	\be
	H'(r)=0 \ .
	\ee
	For a spherically symmetric metric and particularly for our case this function   simplifies to
	\begin{align}
		H(r)=\frac{\sqrt{f(r)}}{r \sin{\theta}}\to \frac{\sqrt{f_3(r)}}{r }\ .
	\end{align}
	The advantage of this approach is that it implicitly incorporates the condition $V(r)=0$, thereby avoiding having to deal with the conserved quantities $E$ and $L$.
	
	\begin{figure*}[!htbp]
		\centering
		\begin{tabular}{@{\hspace*{-0.05\linewidth}}p{0.49\linewidth}@{\hspace*{0.05\linewidth}}p{0.45\linewidth}@{}}
			\centering
			\subfigimg[scale=0.75]{(a)}{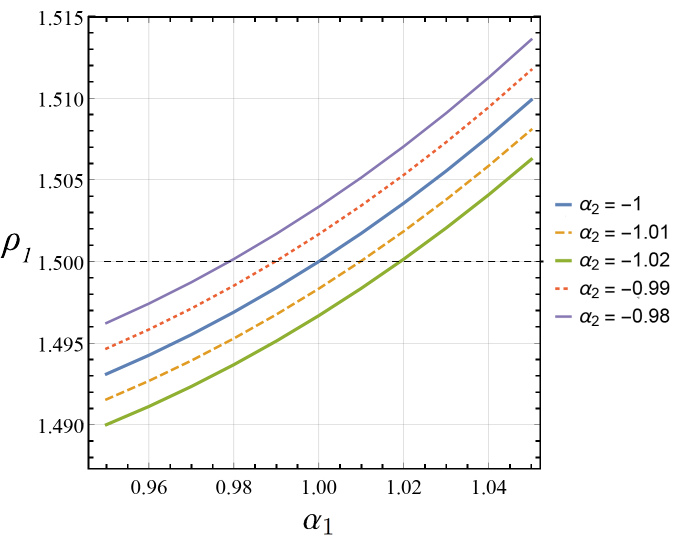} &
			\subfigimg[scale=0.75]{(b)}{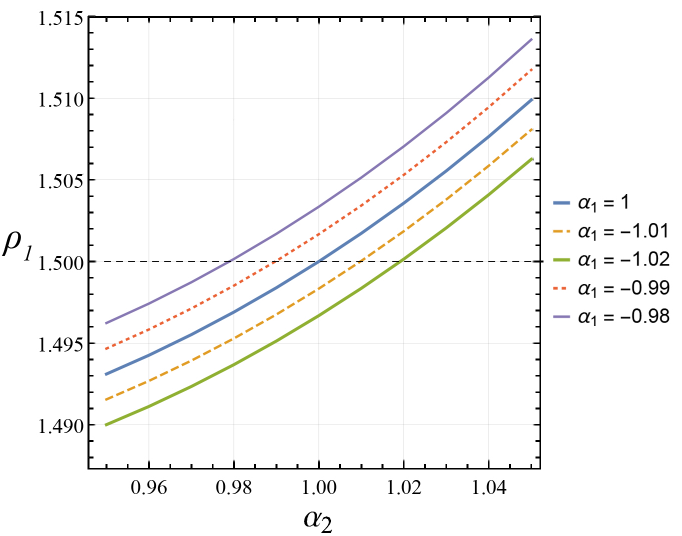}
		\end{tabular}
		\caption{{The location of the outer LR, denoted as $\rho_1$, is shown for different values of the parameters $\alpha_1$ and $\alpha_2$. The solid blue line corresponds to the Schwarzschild case, where $\alpha_1 = 1$ and $\alpha_2 = -1$. (a) We fix $\alpha_2 = -1$, $\beta_1 = 1$, and $r_g = 1$, while varying $\alpha_1$ around its Schwarzschild value, $\alpha_1 = 1$; (b) We fix $\alpha_1 = 1$, $\beta_1 = 1$, and $r_g = 1$, allowing $\alpha_2$ to vary around its Schwarzschild value, $\alpha_2 = -1$.}} \label{fig:LRs}
	\end{figure*}

	The analytic expressions are cumbersome even in this simple setting, so we focus on identifying the LR locations for small deviations from the Schwarzschild black hole parameters, $\delta_1\defeq \alpha_1-1$, $\delta_2\defeq\alpha_2+1$, and $\delta_\beta\defeq \beta_1-1$. Hence
	\be
	H_3=\frac{\sqrt{\rho-1}}{r_g \rho^{3/2}} \left(1+\frac{2\rho-1}{2\rho}\delta_1+\frac{\rho-1}{2\rho}\delta_2\right)+\cO\big(\delta_1^2,\delta_2^2,\delta_\beta^2\big).
	\ee
	As expected, the effect of the deviation of the  ADM mass from its Schwarzschild value $2M=r_g$ is suppressed.
	
	In this approximation the equation $H_3'=0$ has two solutions. The first,
	\be
	\rho_1=\frac{3}{2}+\frac{1}{6}(\delta_1+\delta_2), \label{rho1}
	\ee
	modifies the location of the Schwarzschild LR at $r=\tfrac{3}{2}r_g$, and the second,
	\be
	\rho_2=\frac{5}{6}(\delta_1+\delta_2)\ ,
	\ee
	exists ($i.e. \rho_2>0$) for sufficiently strong violation of the energy conditions. However, for small deviations from the Schwarzschild geometry this LR is invisible for distant observers. Fig.~\ref{fig:LRs} illustrates location of the outer LR for various deviations from the Schwarzschild geometry.

	Some physical intuition can be obtained by comparing the quantitative behavior of the LRs and the energy density near the horizon,
	\begin{align}
		\varrho=-\frac{\delta_1}{8 \pi r_g^2} - \frac{\delta_2x}{4 \pi r_g^3}+\mathcal{O}(x^2).
	\end{align}
	(More details about the {energy-momentum tensor (EMT)} and the energy conditions can be found in Appendix~\ref{appSc}. Its comparison with the above analytical and numerical results indicates that violations of the energy conditions, and, specifically, presence of negative energy density, is correlated with LRs being further from a black hole.
	
	We evaluate the reliability of these estimates by calculating $\rho_1$ using the parameters of the Reissner--Nordström and the Bardeen solutions, and comparing with the exact results.  In the former case, the exact location of the LR,
	\be
	\rho_\mathrm{RN}=\frac{3}{2}+\frac{q^2}{6r_g^2}+\cO\big(q^4\big)
	\ee
	coincides with $\rho_\mathrm{RN3}$ at order $\mathcal{O}(q^2)$  that is obtained from Eq.~\eqref{rho1} with the values of $\alpha_1$ and $\alpha_2$ being set to the values given by Eq.~\eqref{rnlimit}.
	
	{In the latter case
		\be
		\rho_\mathrm{B}=\frac{3}{2}+\frac{7q^2}{12r_g^2}+\cO\big(q^4\big) \ ,
		\ee
		is close to
		\be
		\rho_\mathrm{3}=\frac{3}{2}+\frac{3q^2}{4r_g^2}+\cO\big(q^4\big) \ ,
		\ee
		the values of $\alpha_1$ and $\alpha_2$ were set to the values given by Eq.~\eqref{bdlimit1}.
		
		Note that as $f_3$ was not designed to accommodate the far field behavior of the Reissner--Nordström or Bardeen metric, it has poles at $r>r_g$ for all values of $q$ for both metrics. However, as long as $q/r_g\ll 1$ they are far from $\rho\sim \tfrac{3}{2}$ and thus do not affect the determination of the LR location.

		\section{Discussion}
		
		We have demonstrated the utility of Pad\'{e} approximations for studying both quasinormal modes and light rings of black holes. Our approach employs a two-point Pad\'{e} interpolation to construct a smooth (within some domain of validity) metric that accurately captures both near-horizon and asymptotic features of any desired asymptotically flat, static, spherically symmetric black hole geometry.
		
		Using a matrix discretization scheme, we numerically computed the low-lying quasinormal modes for gravitational perturbations in the background of a metric describing small deviations from the Schwarzschild geometry near the horizon. Our method successfully reproduces the known  quasi-normal frequencies in the appropriate limit, as verified against various numerical and semi-analytic approaches. Notably, we found that even small deviations in the leading components of the metric function near the horizon result in significant changes to the low-lying modes compared to their Schwarzschild values.
		
		Using LIGO--VIRGO observational constraints on the $l=2$ gravitational mode as a benchmark, we examined the sensitivity of these modes to changes in the underlying geometry. We also analyzed the light rings of the geometry, comparing their locations in various approximate metrics to their exact counterparts and relating the results to the energy conditions.
		
		Our analysis reveals that vector-type gravitational perturbations are considerably less sensitive to the near-horizon properties of the black hole than scalar perturbations, as characterized by the near-horizon expansion coefficients $\alpha_1$ and $\alpha_2$. In the case of scalar perturbations, deviations of order $10^{-4}$ in the leading coefficient $\alpha_1$ and of order $10^{-5}$ in the subleading coefficient $\alpha_2$ produce variations exceeding 10\% in the $l\in{0,1,2}$ fundamental modes from their Schwarzschild values. In contrast, for vector-type perturbations, these deviations must be two orders of magnitude larger $(\sim10^{-3})$ to induce a comparable effect. Furthermore, we find that the damping time of the $l=2$ mode is far more sensitive to near-horizon deviations than its oscillation frequency. As a by-product, we provide an improved approximation of the Schwarzschild-like metric compared to our previous work.
		
		We also examined the light rings of the interpolating metrics and demonstrated that Pad\'{e} approximants enable explicit calculations of how the light ring locations depend on the parameters describing the near-horizon geometry. For small deviations from the Schwarzschild metric, two light ring solutions emerge: an outer light ring $\rho_1$, which modifies the classic Schwarzschild light ring position, and a second solution $\rho_2$, which is typically invisible to distant observers. The approximation method proves remarkably accurate when compared with exact solutions for both Reissner-Nordström and Bardeen black holes, even when considering only the leading parameters $\alpha_1$ and $\alpha_2$. Notably, violations of energy conditions, particularly the presence of negative energy density near the horizon, correlate with light rings shifting outward, while the influence of far-field metric deviations on light ring positions remains negligible.
		
		Future work will proceed in several directions. From the perspective of Pad\'{e} approximations, the only distinction between a black hole and a star lies in the metric function at the surface $r = r_0$, where $f(r_0) \neq 0$, corresponding to $F_1 < 1$. We plan to explore the applicability of this approximation in characterizing bosonic stars and other ultracompact objects, thus improving the versatility of the fully model-independent approach. Another natural direction is extending these methods to axially symmetric ultracompact objects, while dynamical (even if slowly evolving) systems provide a crucial setting to test the validity of such approximations beyond equilibrium configurations. It is important to find out if  generalizations of simple approximants like $f_3$ still adequately perform relative to their more sophisticated counterparts. Finally, we will explore the implications of this approach for embedding such solutions in de Sitter spacetime, which describes the universe in its current epoch.

		\acknowledgements
		
		SM and IS are supported by an International Macquarie University Research Excellence
		Scholarship. FS is funded by the ARC Discovery Project Grant No. DP200102152. The work of DRT is supported by the ARC Discovery project Grant No. DP210101279 and the Schwinger Foundation.
		
		\appendix
		
		\section{Some properties of Pad\'{e} approximation}\label{app1}

		The coefficients in Eq.~\eqref{mfrac} are given by
		\begin{align}\label{hankel1}
			F_m & =\frac{-H_m^{(-m+1)} H_{m-2}^{(-m+2)}}{H_{m-1}^{(-m+2)} H_{m-1}^{(-m+1)}}\ , \\
			G_m & =\frac{-H_m^{(-m+1)} H_{m-1}^{(-m+1)}}{H_{m-1}^{(-m+2)} H_m^{(-m)}}\ ,
		\end{align}
		defined through the Hankel determinants that are built from the bisequence of the coefficients $\{c\}_{j=-\infty}^\infty$ of the original expansions \eqref{l1}-\eqref{l2}. They are defined as
		\be\label{hankel2}
		\quad H_j^{(m)}:=\begin{array}{|cccc|}
			c_m & c_{m+1} & \cdots & c_{m+k-1} \\
			c_{m+1} & c_{m+2} & \cdots & c_{m+j} \\
			\vdots & \vdots & & \vdots \\
			c_{m+j-1} & c_{m+j} & \cdots & c_{m+2 j-2}
		\end{array}\ ,
		\ee
		with $H_0^{(m)}\equiv1$. The truncation of the continued fraction in \eqref{mfrac} at some finite $m=n$ results in an approximating function which interpolates between the two finite expansions near $z=0$ and $z=\infty$, in the sense that its Taylor series about these points simultaneously matches that of the (possibly unknown) function $f(z)$ up to order $K_0$ and $K_\infty$ in Eqs.~\eqref{l1} and \eqref{l2}, respectively. It is straightforward to then represent the resulting continued fraction as a rational function, if desired.
		
		As both expansions are in terms of $x=r-r_g$, the coefficients $c_j=\alpha_j$ of Eq.~\eqref{exp-close}. while the coefficients $c_{-j}$ need adjustments. For example, the first three terms of the far field expansion are
		\begin{align}
			\frac{C(r)}{r}&=\beta_1\frac{ r_g}{x}+(\beta_2-\beta_1)\frac{r_g^2}{x^2} \nonumber \\
			& +(\beta_1-2\beta_2+\beta_3)\frac{r_g^3}{x^3} +\cO\big(x^{-4}\big) \ .
		\end{align}

		\section{{The semiclassical metric}}\label{appSc}
		
		We construct the semiclassical metric and discuss its regularity properties and constrained form (see \cite{MMT:22} for an extensive review). A general spherically symmetric metric in Schwarzschild coordinates (with areal radius $r$) is given by
		\be
		ds^2=-e^{2h(t,r)}f(t,r)dt^2+f(t,r)^{-1}dr^2+r^2d\Omega_2\ , \label{sgenm}
		\ee
		while using the advanced null coordinate $v$ results in the form
		\be
		ds^2=-e^{2h_+(v,r)}f_+(v,r)dv^2+2e^{h_+(v,r)}dvdr+r^2d\Omega_2\ . \label{m:vr}
		\ee
		The function  is coordinate-independent function $f(t,r)\equiv f_+\big(v(t,r),r\big)$  is conveniently represented via the MSH mass $M\equiv C/2$  as
		\be
		f=1-\frac{C(t,r)}{r}=1-\frac{C_+(v,r)}{r}\equiv\partial_{\mu} r \partial^{\mu} r\ ,
		\ee
		The functions $h$ and $h_+$ play the role of integrating factors in the coordinate transformation
		\be
		dt=e^{-h}(e^{h_+}dv-f^{-1}dr)\ . \label{trvr-transformation}
		\ee
		For example, the Schwarzschild metric corresponds to $h\equiv 0$, $C\equiv r_g=\mathrm{const}$, and $v=t+r_*$, where $r_*$ is the tortoise coordinate \cite{FN:98,HE:73}.
		
		It is convenient to introduce the effective EMT components
		\begin{align}
			\tau{_t} \equiv &e^{-2h} {T}_{tt}, \label{split1} \\
			\qquad {\tau}{^r} \equiv & T^{rr}\ , \label{split2} \\
			\tau {_t^r} \equiv & e^{-h}  {T}{_t^r}\ . \label{split3} 
		\end{align}
		They allow both for a convenient expression of the regularity conditions and simplify the form of three components of the Einstein equations
		\begin{align}
			&\partial_r C = 8 \pi r^2  {\tau}{_t} , \label{eq:Gtt} \\
			&\partial_t C = 8 \pi r^2 e^h  \tau_t^{r}\ , \label{eq:Gtr} \\
			&\partial_r h = 4 \pi r \left(  \tau_t +  \tau^r \right) / f^2\ . \label{eq:Grr}
		\end{align}
		
		It can be shown that dynamical black hole solutions belong to one of the classes that are distinguished by the behavior of the EMT components in $(t,r)$ coordinates on approach to the apparent horizon. The $k=0$ solution
		corresponds has
		\be
		\tau_t,\tau^r,\tau_t^r \to -\Upsilon^2,
		\ee
		as $r\to r_g$ and describes a shrinking black hole (solutions with $\tau_t^r\to +\Upsilon^2$ describe white holes). On the other hand, the EMT components of $k=1$ solutions scale as $\tau_a\propto f^{1}$ as $r\to r_g$. Static solutions, such as Reissner--Nordstr\"{o}m metric, or various models of regular black holes belong to this class. Dynamical regular black hole models, such as Hayward--Frolov or Simpson--Visser models belong to $k=0$ class.

		The dynamical black hole solutions of this class in $(t,r)$ coordinates have the near-horizon expansion for $r\gtrsim r_g$
		\be\nonumber
		C= r_g-4\sqrt{\pi}r_g^{3/2}\Upsilon\sqrt{ x}+\mathcal{O}( x)\ , \quad h=-\frac{1}{2} \ln{\frac{ x}{\xi}}+\mathcal{O}(\sqrt{ x})\ , \label{k0met}
		\ee
		where  $x\defeq r-r_g(t)$, and the function  $\xi(t)$ is determined by the  choice of time variable.	Consistency of the Einstein equations requires that
		\be
		\frac{d r_g}{dt} {\equiv r'_g} =-\Upsilon\sqrt{\pi r_g\xi}\ .      \label{lumin}
		\ee
		The functions $\Upsilon(t)$ and $\xi(t)$ have to be found from other considerations, such as matching with the standard results for Hawking radiation. Note that $f\propto \sqrt{x}$ near the horizon.

		Our main subjects are the static $k=1$ solutions that we now present. When $\tau_t^r=0$, the leading terms of the effective EMT components are
		\be
		\tau_t=Ef+\ldots, \qquad \tau^r=Pf+\ldots\ ,
		\ee
		where we have not yet determined the behaviour of $f$ as a function of the gap $ x$ and have not restricted the allowed (integer or half-integer) powers of $x$ in higher order-terms. At the  horizon, the energy density and pressure are $\varrho(r_g)=E$ and $p(r_g)=P$, respectively. From the definition of the outer apparent horizon, we must have that $E\leqslant(8\pi r_g^2)^{-1}$. Having the standard curvature invariants finite at the apparent horizon requires $E=-P$ and the absence of fractional powers of $\tilde x$ at least up to quadratic order.

		For $E<(8\pi r_g^2)^{-1}$ the metric functions are
		\begin{align}
			&C=r_g+c_1   x+c_2  x^2+\ldots, \\
			&h=h_1  x+\ldots,
		\end{align}
		where $c_1=8\pi r_g^2 E$ (and thus $f=(1-8\pi E r_g^2)/r_g  x +\mathcal{O}( x^2)$),  the coefficient {$h_1=8(e_2+p_2) r_g^2/(1-c_1)^2$} is determined by the coefficients of the  second order terms in the EMT components, etc.
		
		If the black hole has no hair whatsoever, then $h\equiv 0$, while for short hair it is a rapidly vanishing function of  $x$.
		The condition $h\equiv0$ implies $\tau_t+\tau^r\equiv 0$, i.e. $p=-\varrho$.
		
		The first coefficient in the expansion of Eq.~\eqref{exp-close} is then
		\be
		\alpha_1=1-8\pi E r_g^2.
		\ee
		The next coefficient in the expansion of $f$ is
		\begin{align}
			\alpha_2&=\frac{1}{3\, r_g^2 \left(8 \pi  E\, r_g^2-1\right)^3}\bigg[192 \pi ^2 E\,^2 r_g^4 \left(4 \pi  e_2 r_g^4+3\right)\nonumber\\
			&\quad-24 \pi  E\, r_g^2 \left(8 \pi  e_2 r_g^4+3\right)+12 \pi  e_2 r_g^4\nonumber\\
			&\quad-1536 \pi ^3 E\,^3 r_g^6+3\bigg].\label{c}
		\end{align}
		
		The near-horizon metrics are most conveniently written in the $(v,r)$ coordinates. Using the freedom in re-defining the advanced null coordinate we can write it as
		\begin{align}
			&C_+(v)=r_+(v)+w_1(v)(r-r_+)+\cO\big((r-r_+)^2\big) \ , \\
			&h_+(v)=\chi_1(v)(r-r_+)+\cO\big((r-r_+)^2\big) \ ,
		\end{align}
		where $r_+(v)\equiv r_g\big(t(r_+,v)\big)$ and $w_1\leqslant 1$. This description is valid for both static and dynamic configurations. For a dynamical solution that goes through  $k=1$ type at the horizon formation (and disappearance), $w_1\equiv 1$ \cite{MS:23}.
		
		The Hawking temperature on a static spherically-symmetric background as measured by a distant observer is given by
		\be
		{T_\mathrm{H}=\frac{\kappa}{2\pi}=\frac{f'(r_g)}{4\pi}=\frac{\alpha_1}{4\pi r_g},}
		\ee
		where $\kappa$ is the surface gravity.
		
		We also note that in the dynamical case the only  generalisation of the surface gravity  that is consistent with a finite distant time of horizon formation is the Kodama--Hayward surface gravity \cite{MMT:22-surface_gravity,MS:23-thermo},
		\be
		{\kappa_\mathrm{KH}=\frac{1-w_1}{2r_+}.}
		\ee
		
		The EMT that corresponds to static spherically symmetric metrics is diagonal, with density, (radial) pressure, and tangential pressure being
		\begin{align}
			\varrho=-\tensor{T}{^0_0},\qquad p =\tensor{T}{^1_1}, \quad p_\|=\tensor{T}{^2_2}=\tensor{T}{^3_3},
		\end{align}
		respectively, and as stated above we consider only models with $\varrho=-p$. The EMT belongs to the type I in the Hawking--Ellis classification \cite{HE:73,MV:17}, and conditions for it to satisfy the null energy condition (NEC) are easily verifiable:
		\be
		\varrho+p\geqslant 0, \qquad  \varrho+p_\|\geqslant 0. \label{nec-full}
		\ee
		When supplemented by the requirement
		\be
		\varrho\geqslant 0 \ , \label{wec-full}
		\ee
		it ensures that the {weak} energy condition (WEC) is satisfied.

		For the metrics we consider the first of the conditions of Eq.~\eqref{nec-full} is trivially satisfied, while the nonzero EMT components are
		\begin{align}
			& \tensor{T}{^0_0}=\tensor{T}{^1_1}=-\frac{1-r f'(r)-f(r)}{8\pi r^2} \ ,\\
			& \tensor{T}{^2_2}=\tensor{T}{^3_3}=\frac{2 f'(r)+r f''(r)}{16\pi r} \ .
		\end{align}
		
		At the event horizon of black holes whose near horizon geometry is described by Eq.~\eqref{exp-close}, the NEC is satisfied if
		\be
		\alpha_2\geqslant-1 \ , \label{nec2c}
		\ee
		and is violated otherwise. Eq.~\eqref{wec-full} at the event horizon reads
		\be
		\alpha_1\leqslant 1 \ , \label{wechc}
		\ee
		and both must be true for the WEC to be satisfied.
		
		In the vicinity of the event horizon behaviour of the energy density for $x=r-r_g>0$ is given by
		\be
		{\varrho(x)=-\frac{(\alpha_1-1)}{8\pi r_g^2}-\frac{(\alpha_2+1)}{4\pi r_g^3}x+\cO\big(x^2\big).}
		\ee
		
		\section{Matrix method for black hole perturbations}

		Numerous approaches have been developed for computing QNMs, ranging from analytical to numerical methods. These include the eikonal approximation for $l\gg1$ \cite{P:71,G:72}, the WKB approximation for low harmonic numbers \cite{SW:85,IW:87}, and pseudo-spectral methods \cite{J:17}. While analytical solutions exist for certain effective potentials (see \cite{BV:11} for a comprehensive survey), we implement an enhanced version of the matrix method outlined in \cite{LQ:16,LQ:17} to compute quasinormal frequencies associated with \eqref{radial2}. This approach offers superior efficiency and flexibility compared to traditional finite difference methods.
		This is done by expressing the radial component of the wavefunction through a general master equation
		\be
		G(x,\omega) \phi(x)=0\ ,
		\ee
		where $G(x,\omega)$ represents a differential operator and $x$ denotes the previously defined compact radial coordinate. By discretizing this differential equation across $N$ grid points $x\in{x_i,...,x_N}$, the problem is transformed into a homogeneous matrix equation
		\be\label{mateq}
		\bar{M}(\omega)\phi=0\ ,
		\ee
		where $\phi$ represents a column vector containing the function values $\phi(x_i)$ at each grid point $x_i$. The matrix $M$ is constructed using Cramer's rule to compute derivatives such as
		\be
		\phi''(x_1)=\dfrac{\text{det}(M^{(x_1)}_2)}{\text{det}{(M)}}\ .
		\ee
		This uses the matrix $M$ of Taylor series coefficients of $\phi(x)$ at a grid point $x=x_0$. For instance, at $x=x_0$, one has $\phi=M D$ such that
		\be
		\left(\!\!\begin{array}{c}
			\phi\left(\delta x_1\right) \\
			\phi\left(\delta x_2\right) \\
			\phi\left(\delta x_3\right) \\
			\cdots
		\end{array}\!\!\right)=\left(\!\begin{array}{ccc}
			x_1-x_0 & \frac{\left(x_1-x_0\right)^2}{2} & \frac{\left(x_1-x_0\right)^3}{3 !}   \\
			x_2-x_0 & \frac{\left(x_2-x_0\right)^2}{2} & \frac{\left(x_2-x_0\right)^3}{3 !}  \\
			x_3-x_0 & \frac{\left(x_3-x_0\right)^2}{2} & \frac{\left(x_3-x_0\right)^3}{3 !}   \\
			\ldots & \ldots & \ldots
		\end{array}\!\right)\!\!\left(\!\!\begin{array}{c}
			\phi^{\prime}\left(x_0\right) \\
			\phi^{\prime \prime}\left(x_0\right) \\
			\phi^{\prime \prime \prime}\left(x_0\right) \\
			\cdots
		\end{array}\!\!\right)\nonumber
		\ee
		The determinants are evaluated through $ij$-minor computation followed by cofactor expansion. Non-trivial solutions to \eqref{mateq} exist when:
		\be\label{det}
		\text{det}(\bar{M}(\omega))=0\ .
		\ee
		This yields a polynomial in $\omega$ which can be solved numerically.
		This matrix method has demonstrated its versatility across various applications, including studies of scalar QNMs in asymptotically flat and (anti)-de Sitter Schwarzschild black holes \cite{LQ:17}, Kerr, and Kerr-Sen black holes \cite{LQ:17b}, and accreting Vaidya black holes with specific mass functions \cite{LS:21}, and was used in our previous work \cite{ST:24}. Its advantages over other numerical and semi-analytic approaches are significant: it enables non-uniform grid discretization for enhanced resolution in critical regions, and boundary conditions can be straightforwardly implemented by setting $\bar{M}_{ij}\rightarrow\delta_{ij} \text{ for }i={1,N}$. We specifically avoid double-null construction methods since dynamical backgrounds typically lack analytic double-null foliations.
		
		The method's numerical accuracy is readily demonstrated. Using a uniform grid $x_i\in{0,\tfrac{1}{4},\tfrac{2}{4},\tfrac{3}{4},1}$, we obtain the following $l=m=0$ fundamental mode for the Schwarzschild black hole:
		\be\label{comp1}
		\omega_{000}=0.21863 - 0.19826 I\ .
		\ee
		This can be compared to the established frequency from Leaver's method and pseudo-spectral/AIM approaches \cite{M:22}
		\be
		\omega_{000}=0.22091-0.20979I\ .
		\ee
		Increasing the grid resolution to $x_i\in{0,\tfrac{1}{20},...,\tfrac{19}{20},1}$ yields:
		\be\label{comp2}
		\omega_{000}=0.22088 - 0.20979 I\ .
		\ee
		Our implementation incorporates a crucial modification necessary for the analysis. The condition \eqref{det}, when solved numerically for the quasinormal frequency $\omega$, produces a degree-$N$ polynomial with $N$ complex solutions. Only one of these represents the true quasinormal frequency; the others are spurious solutions whose number increases linearly with grid precision. This poses a significant challenge in identifying the correct frequency as grid resolution increases. While known frequencies can be used to seed root-finding algorithms for \eqref{det}, we lack such prior knowledge for generic near-horizon metrics of the form \eqref{schw-exp}.
		To address this, we use a sequential approach. Given that our interpolating metric \eqref{fS3} and radial wavefunction \eqref{radial2} smoothly approach the Schwarzschild case as $\alpha_1\rightarrow 1$, $\alpha_2\rightarrow -1$, and $\beta_1\rightarrow 1$, we begin from this limit and incrementally adjust the coefficients in finite steps $\{\Delta \alpha_1,\Delta a_2,\Delta \beta_1\}$ until the target value is reached. At each increment, the root finder is seeded with the previous step's result, selecting the solution that minimizes the $L2$ distance from the previously computed quasinormal frequency. This enables reliable tracking of the genuine quasinormal frequency as it evolves smoothly from its Schwarzschild value in concert with the interpolating metric.

		\section{Parameters of the Pad\'{e} approximants} \label{a-detail}
		Here we present parameters of various $n=3$ Pad\'{e} approximates that were used in the text.
		
		For the approximant $f_3$ of Eq.~\eqref{fS3}, parameters of the rational fraction Eq.~\eqref{f3schw} are given by
		\begin{align*}
			A_3&=2 \alpha_1 \beta_1+\alpha_2 \beta_1^2-1\\
			A_2&=3 r_g-\alpha_1 \beta_1^2 r_g-6 \alpha_1 \beta_1 r_g-\alpha_2 \beta_1^3 r_g-3 \alpha_2 \beta_1^2 r_g\nonumber\\
			A_1&=2 \alpha_1 \beta_1^2 r_g^2-\alpha_1 \beta_1^3 r_g^2+6 \alpha_1 \beta_1 r_g^2+2 \alpha_2 \beta_1^3 r_g^2\nonumber\\
			&\quad+3 \alpha_2 \beta_1^2 r_g^2-3 r_g^2\nonumber\\
			A_0&=\alpha_1 \beta_1^3 r_g^3-\alpha_1 \beta_1^2 r_g^3-2 \alpha_1 \beta_1 r_g^3-\alpha_2 \beta_1^3 r_g^3-\alpha_2 \beta_1^2 r_g^3+r_g^3\nonumber\\
			B_2&=\alpha_1 \beta_1^2 r_g-6 \alpha_1 \beta_1 r_g-3 \alpha_2 \beta_1^2 r_g-\beta_1 r_g+3 r_g\nonumber\\
			B_1&=6 \alpha_1 \beta_1 r_g^2-2 \alpha_1 \beta_1^2 r_g^2+3 \alpha_2 \beta_1^2 r_g^2-\beta_1^2 r_g^2+2 \beta_1 r_g^2-3 r_g^2\nonumber\\
			B_0&=\alpha_1 \beta_1^2 r_g^3-2 \alpha_1 \beta_1 r_g^3-\alpha_2 \beta_1^2 r_g^3-\beta_1^3 r_g^3+\beta_1^2 r_g^3\nonumber\\
			&\quad-\beta_1 r_g^3+r_g^3 \ , \nonumber
		\end{align*}
		and of the truncated continued fraction of Eq.~\eqref{f3schw} by
		\begin{align}\label{coeffs2}
			& F_2= \frac{\alpha_1 \beta_1-1}{\beta_1 r_g} \ , \\
			& F_3= \frac{1-2 \alpha_1 \beta_1-\alpha_2 \beta_1^2}{\beta_1 r_g (\alpha_1 \beta_1-1)} \nonumber\\
			& G_1= \frac{1}{\beta_1 r_g} \nonumber\\
			& G_2= \frac{1-\alpha_1 \beta_1}{\beta_1 r_g}\nonumber \\
			&G_3= \frac{2 \alpha_1 \beta_1+\alpha_2 \beta_1^2-1}{\beta_1 r_g (\alpha_1 \beta_1-1)} \nonumber
		\end{align}

		The approximant  $f_\mathrm{RN3}$ of Sec.~\ref{rn-pade} is given as a rational fraction by
		\begin{align*}
			A_3&=\alpha_1^2 \beta_2-2 \alpha_1 \beta_1-\alpha_2 \beta_1^2+\alpha_2 \beta_2+1\\
			A_2&=-\alpha_1^2 \beta_1 \beta_2-3 \alpha_1^2 \beta_2 r_g+\alpha_1 \beta_1^2+6 \alpha_1 \beta_1 r_g-\alpha_1 \beta_2\nonumber\\
			&\quad+3 \alpha_2 \beta_1^2 r_g+\alpha_2 \beta_1^3-2 \alpha_2 \beta_1 \beta_2-3 \alpha_2 \beta_2 r_g-3 r_g\\
			A_1&=2 \alpha_1^2 \beta_1 \beta_2 r_g+\alpha_1^2 \beta_2^2+3 \alpha_1^2 \beta_2 r_g^2-2 \alpha_1 \beta_1^2 r_g+\alpha_1 \beta_1^3\nonumber\\
			&\quad-2 \alpha_1 \beta_1 \beta_2-6 \alpha_1 \beta_1 r_g^2+2 \alpha_1 \beta_2 r_g-3 \alpha_2 \beta_1^2 r_g^2\nonumber\\
			&\quad-2 \alpha_2 \beta_1^3 r_g+4 \alpha_2 \beta_1 \beta_2 r_g+3 \alpha_2 \beta_2 r_g^2+3 r_g^2\\
			A_0&=-\alpha_1^2 \beta_1 \beta_2 r_g^2-\alpha_1^2 \beta_2^2 r_g-\alpha_1^2 \beta_2 r_g^3+\alpha_1 \beta_1^2 r_g^2-\alpha_1 \beta_1^3 r_g\nonumber\\
			&\quad+2 \alpha_1 \beta_1 \beta_2 r_g+2 \alpha_1 \beta_1 r_g^3-\alpha_1 \beta_2 r_g^2+\alpha_2 \beta_1^3 r_g^2\nonumber
				\end{align*}
				
				\begin{align*}
			&\quad +\alpha_2 \beta_1^2 r_g^3-2 \alpha_2 \beta_1 \beta_2 r_g^2-\alpha_2 \beta_2 r_g^3-r_g^3\\
			B_2&=-3 \alpha_1^2 \beta_2 r_g-\alpha_1 \beta_1^2+6 \alpha_1 \beta_1 r_g-\alpha_1 \beta_2+3 \alpha_2 \beta_1^2 r_g\nonumber\\
			&\quad -\alpha_2 \beta_1 \beta_2-3 \alpha_2 \beta_2 r_g+\beta_1-3 r_g\\
			B_1&=3 \alpha_1^2 \beta_2 r_g^2+2 \alpha_1 \beta_1^2 r_g-\alpha_1 \beta_1 \beta_2-6 \alpha_1 \beta_1 r_g^2\\
			&\quad +2 \alpha_1 \beta_2 r_g-3 \alpha_2 \beta_1^2 r_g^2+2 \alpha_2 \beta_1 \beta_2 r_g-\alpha_2 \beta_2^2\nonumber\\
			&\quad+3 \alpha_2 \beta_2 r_g^2+\beta_1^2-2 \beta_1 r_g-\beta_2+3 r_g^2\nonumber\\
			B_0&=-\alpha_1^2 \beta_2 r_g^3-\alpha_1 \beta_1^2 r_g^2+\alpha_1 \beta_1 \beta_2 r_g+2 \alpha_1 \beta_1 r_g^3\\
			&\quad +\alpha_1 \beta_2^2-\alpha_1 \beta_2 r_g^2+\alpha_2 \beta_1^2 r_g^3-\alpha_2 \beta_1 \beta_2 r_g^2+\alpha_2 \beta_2^2 r_g\nonumber\\
			&\quad -\alpha_2 \beta_2 r_g^3-\beta_1^2 r_g+\beta_1^3-2 \beta_1 \beta_2+\beta_1 r_g^2+\beta_2 r_g-r_g^3\nonumber
		\end{align*}
		
		and of the truncated continued fraction of Eq.~\eqref{f3schw} by
		
		\begin{align}
			& F_3= \frac{r_g\beta_1(1-\alpha_1^2 \beta_2-2 \alpha_1 \beta_1-\alpha_2 \beta_1^2-\alpha_2 \beta_2+1)}{(\alpha_1 \beta_1-1) \left(\beta_1^2 r_g^2+\beta_2 r_g^2\right)} \\
			& G_2= \frac{\beta_1 r_g (1-\alpha_1 \beta_1)}{\beta_1^2 r_g^2+\beta_2 r_g^2} \nonumber\\
			&G_3= \frac{\left(\beta_1^2 r_g^2+\beta_2 r_g^2\right) \left(1-\alpha_1^2 \beta_2-2 \alpha_1 \beta_1-\alpha_2 \beta_1^2-\alpha_2 \beta_2\right)}{(1-\alpha_1 \beta_1) \left(\alpha_1 \beta_2^2 r_g^3+\beta_1^3 r_g^3+2 \beta_1 \beta_2 r_g^3\right)} \nonumber
		\end{align}
		where $G_1$ and $F_2$ are still given by \eqref{coeffs2}.
		
		\begin{widetext}
			
			\section{The radial wave equation}\label{appB}
			\noindent For convenience we define $\Theta(x)\equiv \left(\beta_1 x^2 (2 \alpha_1 x+x-1)-x^3+\beta_1^3 (x-1)^3+\beta_1^2 x (x (\alpha_1+x (-\alpha_1+\alpha_2-1)+2)-1)\right)$.
			The radial wave equation \eqref{radial2} is defined by the coefficients
			
			\begin{align}
				\sigma(x)&=\dfrac{1}{\alpha_1^4 r_g^9 \Theta^3 (x-1)^4 x^2 (r_g-r_g x)}\left[(x-1)^2 (1-2 x) \alpha_1^2 \left(\alpha_1 (x (\beta_1-2)-\beta_1) \beta_1 (x+(x-1) \beta_1)\right.\right.\\
				&\quad\left.\left.+x \left(\alpha_2 \beta_1^3-x \alpha_2 (\beta_1+1) \beta_1^2+x\right)\right) \left((x-1) \left(-2 x^6+(x (x (8 \alpha_1+2)-3)+1) \beta_1 x^4\right.\right.\right.\nonumber\\
				&\quad\left.\left.\left.-2 \left(x \left(2 \alpha_1+x \left(-4 \alpha_1+x \left(4 \alpha_1^2+2 \alpha_1-2 \alpha_2+1\right)-3\right)+3\right)-1\right) \beta_1^2 x^3\right.\right.\right.\nonumber\\
				&\quad\left.\left.\left.+2 \left(-\alpha_2^2 x^4-(x-1) (2 (x+1) \alpha_1 x+x-1) \alpha_2 x+(x-1)^2 \alpha_1 (x ((3 x+2) \alpha_1+3)-3)\right) \beta_1^4 x^2\right.\right.\right.\nonumber\\
				&\quad\left.\left.\left.+\left(2 \left(-4 \alpha_1 x^2+x-1\right) \alpha_2 x^2+(x-1) (x (6 \alpha_1+x (2 x-2 \alpha_1 (2 \alpha_1+3)-7)+8)-3)\right) \beta_1^3 x^2\right.\right.\right.\nonumber\\
				&\quad\left.\left.\left.-(x-1) \left((2 x+1) \alpha_2^2 x^3-2 (x-1) ((2 x+1) \alpha_1 x+x-1) \alpha_2 x+(x-1)^2 \alpha_1 (x (\alpha_1+2 x (\alpha_1+2)-6)+2)\right) \beta_1^5 x\right.\right.\right.\nonumber\\
				&\quad\left.\left.\left.+(x-1)^5 ((2 x-1) \alpha_1-2 x \alpha_2) \beta_1^6\right) \alpha_1^2+2 i r_g \left(\alpha_1 \left((x-1)^2+(x-3) x \alpha_1\right)-x \alpha_2\right) \left(x^3-(2 \alpha_1 x+x-1) \beta_1 x^2\right.\right.\right.\nonumber\\
				&\quad\left.\left.\left.+\left((\alpha_1-\alpha_2+1) x^2-(\alpha_1+2) x+1\right) \beta_1^2 x-(x-1)^3 \beta_1^3\right) \left(\alpha_1 \beta_1^3+x (-2 \beta_1 \alpha_1+\alpha_1+\alpha_2 \beta_1) \beta_1^2\right.\right.\right.\nonumber\\
				&\quad\left.\left.\left.+x^2 (\beta_1 (\beta_1+1) (\alpha_1 (\beta_1-2)-\alpha_2 \beta_1)+1)\right) \omega \right) r_g^{10}+\Theta (6 (x-1) x+2) (r_g-r_g x)^3 \alpha_1^4 \left(\alpha_1 (x (\beta_1-2)\right.\right.\nonumber\\
				&\quad\left.\left.-\beta_1) \beta_1 (x+(x-1) \beta_1)+x \left(\alpha_2 \beta_1^3-x \alpha_2 (\beta_1+1) \beta_1^2+x\right)\right)^2 r_g^7\right.\nonumber\\
				&\quad\left.-(1-x) \left(i (\alpha_1-x \alpha_1)^2 \left(\alpha_1 (x-1)^2+(x-3) x \alpha_1^2-x \alpha_2\right) \beta_1 \left(\alpha_1 (x (\beta_1-2)-\beta_1) \beta_1 (x+(x-1) \beta_1)\right.\right.\right.\nonumber\\
				&\quad\left.\left.\left.+x \left(\alpha_2 \beta_1^3-x \alpha_2 (\beta_1+1) \beta_1^2+x\right)\right) \left[x^2 \alpha_1^2 \beta_1^2 (\beta_1-x (\beta_1-2))^2+\alpha_1 \beta_1 \left(-4 x^4+6 (x-1) \beta_1 x^3\right.\right.\right.\right.\nonumber\\
				&\quad\left.\left.\left.\left.+2 (x (x (2 \alpha_2-3)+6)-3) \beta_1^2 x^2-2 (x-1) (x (\alpha_2 x+x-2)+1) \beta_1^3 x+(x-1)^4 \beta_1^4\right)+x \left(x^3 \alpha_2^2 \beta_1^4\right.\right.\right.\right.\nonumber\\
				&\quad\left.\left.\left.\left.-2 \alpha_2 \left(x^3-(x-1) \beta_1 x^2+(x-1)^2 \beta_1^2 x+(x-1)^3 \beta_1^3\right) \beta_1^2+x \left(x^2-2 (x-1) \beta_1 x+3 (x-1)^2 \beta_1^2\right)\right)\right] \omega  r_g^9\right.\right.\nonumber\\
				&\quad\left.\left.-\left(x^3-(2 \alpha_1 x+x-1) \beta_1 x^2+\left((\alpha_1-\alpha_2+1) x^2-(\alpha_1+2) x+1\right) \beta_1^2 x-(x-1)^3 \beta_1^3\right) \left(\alpha_1 (x (\beta_1-2)\right.\right.\right.\nonumber\\
				&\quad\left.\left.\left.-\beta_1) \beta_1 (x+(x-1) \beta_1)+x \left(\alpha_2 \beta_1^3-x \alpha_2 (\beta_1+1) \beta_1^2+x\right)\right)^2 \omega  \left(-i (x-1)^2 \left(x^2 (\alpha_1+1)-1\right) \alpha_1^3\right.\right.\right.\nonumber\\
				&\quad\left.\left.\left.-r_g \left(\alpha_1 \left((x-1)^2+(x-3) x \alpha_1\right)-x \alpha_2\right)^2 \omega \right) r_g^9+\alpha_1^4 \left(\Theta^3 \omega ^2 r_g^{10}+(x-1)^2 x \left(\alpha_1 (x (\beta_1-2)-\beta_1) \beta_1 (x+(x-1) \beta_1)\right.\right.\right.\right.\nonumber\\
				&\quad\left.\left.\left.\left.+x \left(\alpha_2 \beta_1^3-x \alpha_2 (\beta_1+1) \beta_1^2+x\right)\right) \left(L (L+1) \left(-x^3+(2 \alpha_1 x+x-1) \beta_1 x^2\right.\right.\right.\right.\right.\nonumber\\
				&\quad\left.\left.\left.\left.\left.+(x (\alpha_1+x (-\alpha_1+\alpha_2-1)+2)-1) \beta_1^2 x+(x-1)^3 \beta_1^3\right)^2+3 \left((x-1) (-x \alpha_1+\alpha_1+2 x \alpha_2) \beta_1^3\right.\right.\right.\right.\right.\nonumber\\
				&\quad\left.\left.\left.\left.\left.+x (2 (x-1) \alpha_1+3 x \alpha_2) \beta_1^2+6 x^2 \alpha_1 \beta_1-3 x^2\right) \left(x^3-(2 \alpha_1 x+x-1) \beta_1 x^2+\left((\alpha_1-\alpha_2+1) x^2\right.\right.\right.\right.\right.\right.\nonumber\\
				&\quad\left.\left.\left.\left.\left.\left.-(\alpha_1+2) x+1\right) \beta_1^2 x-(x-1)^3 \beta_1^3\right)+3 x \left(\alpha_1 (x (\beta_1-2)-\beta_1) \beta_1 (x+(x-1) \beta_1)\right.\right.\right.\right.\right.\nonumber\\
				&\quad\left.\left.\left.\left.\left.+x \left(\alpha_2 \beta_1^3-x \alpha_2 (\beta_1+1) \beta_1^2+x\right)\right) \left((\beta_1 (2 \alpha_1 (\beta_1-3)-3 \alpha_2 \beta_1+\beta_1-2)+3) x^2\right.\right.\right.\right.\right.\nonumber\\
				&\quad\left.\left.\left.\left.\left. -2 \beta_1 (\alpha_1 \beta_1+\beta_1-1) x+\beta_1^2\right)\right) r_g^8\right)\right) r_g^2\right]\nonumber
			\end{align}
			
			\vspace{-0.5cm}
			\begin{align}
				\tau(x)&=\dfrac{r_g(1-x)}{\alpha_1^2 \Theta^3 (x-1)^2 x (r_g-r_g x)}\bigg[\alpha_1 \beta_1 ((\beta_1-2) x-\beta_1) (\beta_1 (x-1)+x)+x \left(\alpha_2 \beta_1^3-\alpha_2 (\beta_1+1) \beta_1^2 x+x\right)\bigg]\\
				&\times \Big[\alpha_1^2 (x-1)^2 \left(\alpha_1 \beta_1^6+\beta_1^3 x^2 \left(\beta_1 \left(-3 \alpha_1^2 \beta_1+2 \alpha_1 (\beta_1 (7 \beta_1-5)+3)-2 \alpha_2 (\beta_1-1) \beta_1\right)-1\right)\right.\nonumber\\
				&\quad\left.+2 \beta_1^3 x^3 \left(\alpha_1^2 \beta_1 (4 \beta_1-5)+\alpha_1 (3-\beta_1 (\beta_1 (3 \alpha_2+8 \beta_1-9)+6))+\alpha_2 \beta_1 (\beta_1 (3 \beta_1-2)+1)\right)\right.\nonumber\\
				&\quad\left.+\beta_1 x^4 \left(\beta_1 \left(\beta_1 \left(\alpha_1^2 (-(\beta_1-2)) (7 \beta_1-2)+\alpha_1 (\beta_1 (2 \alpha_2 (5 \beta_1-4)+\beta_1 (9 \beta_1-14)+6)-6)\right.\right.\right.\right.\nonumber\\
				&\quad\left.\left.\left.\left.+\alpha_2 (2-\beta_1 (\beta_1 (3 \alpha_2+6 \beta_1-2)+2))+3\right)-2\right)+1\right)-2 x^5 (\beta_1 (\beta_1+1) (\alpha_1 (\beta_1-2)-\alpha_2 \beta_1)+1) (\beta_1 (-\alpha_1 (\beta_1-2)\right.\nonumber\\
				&\quad\left.+\beta_1 (\alpha_2+\beta_1-1)+1)-1)+2 \alpha_1 (1-3 \beta_1) \beta_1^5 x\right)+2 i r_g \omega  \left(\alpha_1 \left(\alpha_1 (x-3) x+(x-1)^2\right)-\alpha_2 x\right) \left(\beta_1^2 x \left(x^2 (\alpha_1-\alpha_2+1)\right.\right.\nonumber\\
				&\quad\left.\left.-(\alpha_1+2) x+1\right)-\beta_1 x^2 (2 \alpha_1 x+x-1)+x^3-\beta_1^3 (x-1)^3\right) \left(\alpha_1 \beta_1^3+x^2 (\beta_1 (\beta_1+1) (\alpha_1 (\beta_1-2)-\alpha_2 \beta_1)+1)\right.\nonumber\\
				&\quad\left.+\beta_1^2 x (-2 \alpha_1 \beta_1+\alpha_1+\alpha_2 \beta_1)\right)\Big]\nonumber
			\end{align}
					
			\vspace{-0.5cm}
				\begin{align}
				\gamma(x)&=-\Theta^2\left(\alpha_1 \beta_1^3+\beta_1^2 x (-2 \alpha_1 \beta_1+\alpha_1+\alpha_2 \beta_1)+x^2 (\beta_1 (\beta_1+1) (\alpha_1 (\beta_1-2)-\alpha_2 \beta_1)+1)\right)^2\hspace{4cm}
			\end{align}

		\end{widetext}

	\end{document}